\documentclass{tlp}

\begin{document}

\newcommand{\hsp}{\mbox{$\hspace*{5mm}$}}
\newcommand{\hspp}{\mbox{$\hspace*{35mm}$}}

\newtheorem{defn}{Definition}[section]
\newtheorem{thm}{Theorem}[section]
\newtheorem{lemma}{Lemma}[section]
\newtheorem{ex}{Example}[section]
\newtheorem{prop}{Proposition}[section]
\newtheorem{fact}{Fact}[section]
\newtheorem{coro}{Corollary}[section]
\newtheorem{obs}{Observation}[section]
\newtheorem{alg}{Algorithm}[section]

\newcommand{\ra}{\mbox{$\rightarrow$}}
\newcommand{\la}{\mbox{$\leftarrow$}}
\newcommand{\rimplies}{\mbox{$\Rightarrow$}}
\newcommand{\limplies}{\mbox{$\Leftarrow$}}

\newcommand{\cf}{\mbox{{\sl{conf\,}}}}    

\newcommand{\conf}{\mbox{$\langle[\alpha,\beta],~ [\gamma,\delta]\rangle$}}
\newcommand{\negconf}{\mbox{$\langle\rangle$}}

\newcommand{\confi}[1]{\mbox{$\langle[\alpha_{#1},\beta_{#1}],~ [\gamma_{#1},\delta_{#1}]\rangle$}}
\newcommand{\tcond}{\mbox{$\alpha_1 \leq \alpha_2 \wedge \beta_1 \leq \beta_2 \wedge \gamma_2 \leq \gamma_1 \wedge \delta_2 \leq \delta_1$}}
\newcommand{\kcond}{\mbox{$\alpha_1 \leq \alpha_2 \wedge \beta_1 \leq \beta_2 \wedge \gamma_1 \leq \gamma_2 \wedge \delta_1 \leq \delta_2$}}
\newcommand{\pcond}{\mbox{$\alpha_2 \leq \alpha_1 \wedge \beta_1 \leq \beta_2 \wedge \gamma_2 \leq \gamma_1 \wedge \delta_1 \leq \delta_2$}}
\newcommand{\lang}{\mbox{${\cal L}$}}
\newcommand{\plang}{\mbox{${\cal L}_p$}}

\newcommand{\muone}{\mbox {$\mu_1$}}
\newcommand{\muk}{\mbox {$\mu_k$}}
\newcommand{\crone}{\mbox {$c_{r_1}$}}
\newcommand{\crk}{\mbox {$c_{r_k}$}}

\newcommand{\join}{\mbox {$\oplus$}}
\newcommand{\meet}{\mbox {$\otimes$}}
\newcommand{\joint}{\mbox {$\oplus_t$}}
\newcommand{\meett}{\mbox {$\otimes_t$}}
\newcommand{\joink}{\mbox {$\oplus_k$}}
\newcommand{\meetk}{\mbox {$\otimes_k$}}
\newcommand{\joinp}{\mbox {$\oplus_p$}}
\newcommand{\meetp}{\mbox {$\otimes_p$}}

\newcommand{\orA}{\mbox {$\vee_{\mu_A}$}}
\newcommand{\orpc}{\mbox {$\vee_{pc}$}}
\newcommand{\andm}{\mbox {$\wedge_m$}}
\newcommand{\andind}{\mbox {$\wedge_{ind}$}}
\newcommand{\andig}{\mbox {$\wedge_{ig}$}}
\newcommand{\andpc}{\mbox {$\wedge_{pc}$}}

\newcommand{\andmode}{\mbox {$\wedge_{\mu_r}$}}

\newcommand{\ormode}{\mbox {$\vee_{\mu_p}$}}

\newcommand{\worldij}{\mbox {$W_{ij}$}}
\newcommand{\worldi}{\mbox {$W_{i}$}}

\newcommand{\worldzz}{\mbox {$W_{00}$}}
\newcommand{\worldzb}{\mbox {$W_{0\bot}$}}
\newcommand{\worldzo}{\mbox {$W_{01}$}}
\newcommand{\worldbz}{\mbox {$W_{\bot 0}$}}
\newcommand{\worldbb}{\mbox {$W_{\bot \bot}$}}
\newcommand{\worldbo}{\mbox {$W_{\bot 1}$}}
\newcommand{\worldoz}{\mbox {$W_{10}$}}
\newcommand{\worldob}{\mbox {$W_{1\bot}$}}
\newcommand{\worldoo}{\mbox {$W_{11}$}}

\newcommand{\wij}{\mbox {$w_{ij}$}}
\newcommand{\wi}{\mbox {$w_{i}$}}

\newcommand{\wzz}{\mbox {$w_{00}$}}
\newcommand{\wzb}{\mbox {$w_{0\bot}$}}
\newcommand{\wzo}{\mbox {$w_{01}$}}
\newcommand{\wbz}{\mbox {$w_{\bot 0}$}}
\newcommand{\wbb}{\mbox {$w_{\bot \bot}$}}
\newcommand{\wbo}{\mbox {$w_{\bot 1}$}}
\newcommand{\woz}{\mbox {$w_{10}$}}
\newcommand{\wob}{\mbox {$w_{1\bot}$}}
\newcommand{\woo}{\mbox {$w_{11}$}}

\newcommand{\wo}{\mbox {$w_{1}$}}
\newcommand{\wz}{\mbox {$w_{0}$}}
\newcommand{\wb}{\mbox {$w_{\bot}$}}

\newcommand{\conflat}{\mbox {${\cal L}_c$}}

\def\blackbox{{\rule{2.2mm}{2.2mm}}}
\def\qed{\hspace*{\fill}\blackbox}

\newcommand{\lla}{\mbox {$\longleftarrow$}}

\newcommand{\vlla}{\mbox{$<\hspace*{-9pt} \rule[2pt]{80pt}{0.5pt}$ }}

\newcommand{\prule}{\mbox {$\stackrel{\bf \conf}{\vlla}$}}

\newcommand{\prulearg}[1]{\mbox {$\stackrel{\bf {#1}}{\vlla}$}}

\newcommand{\shortprule}{\mbox {$\stackrel{\bf c}{\la}$}}

\newcommand{\shortprulearg}[1]{\mbox {$\stackrel{\bf {#1}}{\la}$}}

\newcommand{\stdrule}{\mbox {$A~~\prule~~B_1, \ldots, B_m$}}

\newcommand{\stdrulearg}[1]{\mbox {$A~~\prulearg{#1}~~B_1, \ldots, B_m$}}

\newcommand{\shortstdrule}{\mbox {$A~~\shortprule~~B_1, \ldots, B_m$}}

\newcommand{\shortstdrulearg}[1]{\mbox {$A~~\shortprulearg{#1}~~B_1, \ldots, B_m$}}

\newcommand{\stdfact}{\mbox {$A~~\prule$}} 

\newcommand{\shortstdfact}{\mbox {$A~~\shortprule$}}

\newcommand{\stdfactarg}[1]{\mbox {$A~~\prulearg{#1}$}}

\newcommand{\shortstdfactarg}[1]{\mbox {$A~~\shortprulearg{#1}$}}

\newcommand{\const}{\mbox {${\cal C}$}}

\newcommand{\ie}{\mbox {\em i.e.~}}

\newcommand{\eg}{\mbox {\em e.g.~}}

\newcommand{\bigeg}{\mbox {\em E.g.~}}

\newcommand{\false}{\mbox {$\langle[0,0],~[1,1]\rangle$}}

\newcommand{\true}{\mbox {$\langle[1,1],~[0,0]\rangle$}}

\newcommand{\tpann}{\mbox {$T_P^{NS}$}}

\newcommand{\prob}{\mbox {${\bf Prob}$}}

\newcommand{\vs}{\vspace{1ex}}              
\newcommand{\vsp}{\vspace{2ex}}             
\newcommand{\vspp}{\vspace{4ex}}            

\newcommand{\negvs}{\vspace{-1ex}}          

\newcommand{\infosystems}{Information Systems}

\newcommand{\tkde}{IEEE Transactions on Knowledge and Data Engineering}

\title{On A Theory of Probabilistic Deductive Databases}


\author[Laks V. S. Lakshmanan and Fereidoon Sadri]
{LAKS V. S. LAKSHMANAN\thanks{Research was supported by grants from
the Natural Sciences and Engineering Research Council of Canada 
and NCE/IRIS.}\\
Department of Computer Science\\
Concordia University\\
Montreal, Canada\\
and\\
K.R. School of Information Technology\\
IIT -- Bombay\\
Mumbai, India\\
\email{laks@cs.concordia.ca}
\and FEREIDOON SADRI\thanks{Research was supported by grants from
NSF and UNCG.}\\
Department of Mathematical Sciences\\
University of North Carolina\\
Greensboro, NC, USA\\
\email{sadri@uncg.edu}}

\date{}
\maketitle

\begin{abstract}
We propose a framework for modeling uncertainty where both belief and doubt 
can be given independent, first-class status. We adopt probability theory 
as the mathematical formalism for manipulating uncertainty. An agent can 
express the uncertainty in her knowledge about a piece of information in the 
form 
of a {\em confidence level}, consisting of a pair of intervals of probability, 
one for each of her belief and doubt. The space of confidence levels naturally 
leads to the notion of a {\em trilattice}, similar in spirit to Fitting's 
bilattices. Intuitively, the points in such a trilattice can be ordered 
according to truth, information, or precision. 
We develop a framework for {\em probabilistic deductive databases}  
by associating confidence levels with the facts and rules of a classical 
deductive database. 
While the trilattice structure offers a variety of 
choices for defining the semantics of probabilistic deductive databases, our 
choice of semantics is based on the truth-ordering, which we find to be 
closest to the classical framework for deductive databases. In addition to 
proposing a declarative semantics based on valuations and an equivalent 
semantics based on fixpoint theory, we also propose a proof procedure and 
prove it sound and complete. We show that while classical Datalog query 
programs have a polynomial time data complexity, certain query programs in the 
probabilistic deductive database framework do not even 
terminate on some input databases. We identify a large natural class of query 
programs of practical interest in our framework, and show that programs 
in this class possess polynomial time data complexity, \ie not only do they 
terminate on every input database, they are guaranteed to do so in a 
number of steps polynomial in the input database size. 

\end{abstract}


\section{Introduction} \label{intro}

Knowledge-base systems must typically deal with 
imperfection in knowledge, in particular, in the form of incompleteness, 
inconsistency, and uncertainty.
With this motivation, several frameworks for 
manipulating data and knowledge have been proposed in the form 
of extensions to classical logic programming and deductive 
databases to cope with imperfections in available knowledge.  
Abiteboul, {\em et al.} \cite{abiteboul-etal-nulls-tcs91}, Liu 
\cite{liu-ddb-nulls-ilps90}, and 
Dong and Lakshmanan \cite{dong-laks-92-jicslp-ddb-nulls} 
dealt with deductive databases with incomplete 
information in the form of null values. Kifer and Lozinskii 
\cite{kifer-loz-logic-incons-lics89,kifer-loz-logic-incons-jar} 
have  developed a logic for reasoning with inconsistency. 
Extensions to logic programming and deductive databases for handling 
uncertainty are numerous.
They can broadly be categorized into non-probabilistic and probabilistic
formalisms.
We review previous work in these fields, with special emphasis on 
probabilistic logic programming, because of its relevance to this paper. 

\vs
\noindent
{\bf Non-probabilistic Formalisms}

\noindent
{\sl (1) Fuzzy logic programming}: This was essentially introduced by 
van Emden in his seminal paper on quantitative deduction 
\cite{vEm86:QuanDedFPTh}, and further developed by various researchers, 
including  Steger {\em et al.} \cite{steger-etal-qddb-icde89}, 
Schmidt {\em et al.} \cite{schmidt-etal-qddb-dood89}. 

\noindent
{\sl (2) Annotated logic programming}: 
This framework was introduced by Subrahmanian \cite{Sub87:QLP}, 
and later studied by Blair and Subrahmanian 
\cite{Bl89:ParaLP,blair-vs-para-lp-tcs89}, and Kifer and 
Li \cite{Ki88:SemESUnc}. While Blair and Subrahmanian's focus was 
paraconsistency, Kifer and Li extended the framework of \cite{Sub87:QLP} 
into providing a formal 
semantics for rule-based systems with uncertainty. Finally, this framework was 
generalized by Kifer and Subrahmanian into the generalized annotated 
programming (GAP) framework  
\cite{KiSub92:GAPs}). All these frameworks are inherently based on a 
lattice-theoretic semantics. 
Annotated logic programming has also been employed with the probabilistic
approach, which we will discuss further below.

\noindent
{\sl (3) Evidence theoretic logic programming}: This has been mainly 
studied by Baldwin and Monk 
\cite{baldwin-monk-evidential-fuzzy-lp-tr87} and 
Baldwin \cite{baldwin-evidential-lp-jfuzzy87}). They use Dempster's evidence 
theory as the basis for dealing with uncertainty in their logic programming 
framework. 

\vs
\noindent
{\bf Probabilistic Formalisms}

\noindent
Indeed, there has been substantial amount of research into 
probabilistic logics ever since Boole 
\cite{boole-prob-logic-bk-1854}. 
Carnap \cite{carnap-prob-logic-bk62} is a seminal work on probabilistic 
logic. Fagin, Halpern, and Megiddo \cite{fagin-etal-prob-logic-inc92} 
study the 
satisfiability of systems of probabilistic constraints from a model-theoretic 
perspective. Gaifman \cite{gaifman-prob-logic-isrjmath64} extends probability 
theory by borrowing 
notions and techniques from logic. Nilsson \cite{nilsson-prob-logic-jai86} 
uses a ``possible worlds" 
approach to give model-theoretic semantics for probabilistic logic. 
Hailperin's \cite{Hailperin84:ProbLog} notion of probabilistic entailment is 
similar to that of Nilsson.
Some of the probabilistic logic programming works are based on 
probabilistic logic approaches, such as Ng and Subrahmanian's
work on probabilistic logic programming \cite{Ng:PLP} and Ng's recent
work on empirical databases \cite{Ng:tkde97}.
We discuss these works further below.
We will not elaborate on probabilistic logics 
any more and refer the reader to Halpern \cite{Hal-fol+prob-jAI-90} for 
additional information. 

Works on probabilistic logic programming and deductive databases
can be categorized into two main approaches, annotation-based,
and implication based.

\noindent
{\bf Annotation Based Approach}: 
Ng and Subrahmanian \cite{Ng:PLP} were the first to propose a probabilistic 
basis for logic programming. Their syntax borrows from that of 
annotated logic programming \cite{KiSub92:GAPs}, although the semantics are 
quite different. The idea is that uncertainty is always associated with 
individual atoms (or their conjunctions and disjunctions), while the rules 
or clauses are always kept classical. 

In \cite{Ng:PLP}, uncertainty in an atom is modeled by 
associating a probabilistic truth value with it, and by asserting that it lies 
in an interval. The main interest is in characterizing how precisely we can 
``bound" the probabilities associated with various atoms. In terms of the 
terminology of belief and doubt, we can say, following Kifer and Li 
\cite{Ki88:SemESUnc}, that the combination of belief and doubt about a 
piece of information might lead to an interval of probabilities, as opposed 
a precise probabilities. But, as pointed out in \cite{Ng:PLP}, even if 
one starts with precise point probabilities for atomic events, probabilities 
associated with compound events can only be calculated to within 
some exact upper and lower bounds, thus naturally necessitating intervals. 
But then, the same argument can be made for an agent's belief as well as 
doubt about a fact, \ie they both could well be intervals. In this 
sense, we can say that the model of \cite{Ng:PLP} captures only the belief. 
A second important characteristic of this model is that it makes a 
conservative assumption that nothing is known about the interdependence 
of events (captured by the atoms in an input database), and thus has 
the advantage of not having to make the often unrealistic independence 
assumption. However,  by being conservative, it makes it impossible to 
take advantage of the (partial) knowledge a user may have about the 
interdependence among some of the events. 

From a technical perspective, only annotation constants are 
allowed in \cite{Ng:PLP}. Intuitively, this means only constant probability 
ranges may be associated with atoms. This was generalized in a subsequent 
paper by Ng and Subrahmanian 
\cite{NgSub90:SFforSS+CP-in-DD} to allow annotation variables and functions. 
They have developed fixpoint and 
model-theoretic semantics, and 
provided a sound and weakly complete proof procedure.  
G\"{u}ntzer {\em et al.} \cite{guntzer-etal-newdir-unc-ddb-sigmod91}
have proposed 
a sound (propositional) probabilistic calculus 
based on conditional probabilities, for reasoning in the presence of 
incomplete information. 
Although they 
make use of a datalog-based interface to implement this calculus, 
their framework is actually propositional. 
In related works, Ng and Subrahmanian have extended their 
basic probabilistic logic programming framework to capture stable 
negation in \cite{NgSub90:SS-for-PLP}, and developed a basis for 
Dempster-Shafer theory in \cite{NgSub:DS-theory-91}. 

\vs
\noindent
{\bf Implication Based Approach}: 
While many of the quantitative deduction frameworks (van Emden 
\cite{vEm86:QuanDedFPTh}, 
Fitting \cite{Fi88:LPLat,fitting-bilattice-lp-jlp91}, 
Debray and Ramakrishnan
\cite{debray-raghu-ghcp-jan94},\footnote{The framework proposed in
\cite{debray-raghu-ghcp-jan94} unifies Horn clause based computations in a
variety of settings, including that of quantitative deduction as
proposed by van Emden \cite{vEm86:QuanDedFPTh}, within one abstract formalism.
However, in view of the assumptions made in \cite{debray-raghu-ghcp-jan94}, 
not all
probabilistic conjunctions and disjunctions are permitted by that formalism.}
etc.) are implication 
based, the first implication based 
framework for probabilistic deductive databases was proposed in 
\cite{laks-sadri-pddb-ilps94}. 
The idea behind implication based approach is to associate uncertainty 
with the facts as well as rules in a deductive database. 
Sadri \cite{Sa:RAQ,Sad-IST-icde-90} in a number of 
papers developed a hybrid method called Information Source Tracking 
(IST) for modeling uncertainty in (relational)
databases which combines symbolic and numeric approaches to modeling 
uncertainty. Lakshmanan and Sadri \cite{LS:DIST,Lakshmanan-Sadri:is97}
pursue the deductive 
extension of this model using the implication based approach. 
Lakshmanan \cite{laks-epistemic-fsttcs94} generalizes the idea behind IST  
to model uncertainty by characterizing the set of (complex) scenarios under 
which certain (derived) events might be believed or doubted given a knowledge 
of the applicable belief and doubt scenarios for basic events. He also 
establishes a connection between this framework and modal logic. 
While both \cite{laks-epistemic-fsttcs94,LS:DIST} are implication based 
approaches, strictly speaking, they do not require any commitment to a 
particular formalism (such a probability theory) for uncertainty 
manipulation. Any formalism that allows for a consistent calculation of 
numeric certainties associated with boolean combination of basic events, 
based on given certainties for basic events, can be used for computing the 
numeric certainties associated with derived atoms.  

Recently, Lakshmanan and Shiri \cite{tkde97} unified and 
generalized all known implication based frameworks for deductive 
databases with uncertainty (including those that use formalisms other than 
probability theory) 
into a more abstract framework called the parametric framework. The notions 
of conjunctions, disjunctions, and certainty propagations (via rules) are 
parameterized and can be chosen based on the applications. Even the 
domain of certainty measures can be chosen as a parameter. Under such broadly 
generic conditions, they proposed a declarative semantics and an equivalent 
fixpoint semantics. They also proposed a sound and complete proof procedure. 
Finally, they characterized conjunctive query containment in this framework 
and provided necessary and sufficient conditions for containment for 
several large classes of query programs. Their results can be applied to 
individual implication based frameworks as the latter can be seen as instances 
of the parametric framework. Conjunctive query containment is one of 
the central problems in query optimization in databases. 
While the framework of this paper can also be realized as an instance of the 
parametric framework, the concerns and results there are
substantially different 
from ours. In particular, to our knowledge, this is the first paper to address 
data complexity in the presence of (probabilistic) uncertainty. 

\newpage
\noindent
{\bf Other Related Work} 

\vs
\noindent
Fitting \cite{Fi88:LPLat,fitting-bilattice-lp-jlp91} has 
developed an elegant framework for quantitative logic programming 
based on bilattices, an algebraic structure 
proposed by Ginsburg \cite{ginsburg-bilattice-compint88} 
in the context of many-valued logic programming. This was the first to 
capture both belief and doubt in one uniform logic programming 
framework.  
In recent work, Lakshmanan {\em et al.} \cite{probview96} 
have proposed a model and algebra for probabilistic relational databases. 
This framework allows the user to choose notions of conjunctions and 
disjunctions based on a family of strategies. In addition to developing 
complexity results, they also address the problem of efficient maintenance of 
materialized views based on their probabilistic relational algebra. One of 
the strengths of their model is not requiring any restrictive independence 
assumptions among the facts in a database, unlike previous work on 
probabilistic relational databases \cite{BGP92}. 
In a more recent work, Dekhtyar and 
Subrahmanian \cite{dekhtyarVS97} developed an annotation based framework 
where the user can have a parameterized notion of conjunction and disjunction. 
In not requiring independence assumptions, and being able to allow the user 
to express her knowledge about event interdependence by means of a parametrized 
family of conjunctions and disjunctions,
both \cite{dekhtyarVS97,probview96} 
have some similarities to this paper. However, chronologically, 
the preliminary version of this paper \cite{laks-sadri-pddb-ilps94} was 
the first to incorporate such an idea in a probabilistic framework. 
Besides, the frameworks of \cite{dekhtyarVS97,probview96} are substantially 
different from ours. 
In a recent work Ng \cite{Ng:tkde97} studies {\em empirical} databases, where
a deductive database is enhanced by empirical clauses representing
statistical information.    
He develops a model-theoretic semantics, and
studies the issues of consistency and query processing in such databases.
His treatment is probabilistic, where probabilities are obtained from
statistical data, rather than being subjective probabilities.
(See Halpern \cite{Hal-fol+prob-jAI-90} for a comprehensive discussion
on statistical and 
subjective probabilities in logics of probability.) 
Ng's query processing algorithm attempts to resolve a query using the 
(regular) deductive component of the database. If it is not successful, 
then it reverts to the empirical component, using the notion of
{\em most specific reference class} usually used in statistical
inferences.
Our framework is quite different in that every rule/fact is
associated with a confidence level (a pair of probabilistic intervals 
representing belief and doubt), which may be subjective,  
or may have been obtained from underlying statistical data. 
The emphasis of our work is 
on ({\em i}) the characterization of different modes
for combining confidences,
({\em ii}) semantics, and, in particular,
({\em iii}) termination and complexity issues.

\vsp
The contributions of this paper are as follows. 
\begin{itemize} 
\item We associate a {\em confidence level} with facts and rules (of a 
deductive database). 
A confidence level comes with both a {\sl belief} and 
a {\sl doubt}\footnote{We 
specifically avoid the term {\em disbelief} because of its possible implication 
that it is the complement of belief, in some sense. In our framework, 
doubt is not necessarily the truth-functional complement of belief.} 
(in what is being asserted) [see Section \ref{motiv} for a motivation]. 
Belief and doubt are subintervals of $[0,1]$ representing probability ranges. 

\item We show that 
confidence levels have an interesting algebraic structure called 
{\sl trilattices} as their basis (Section \ref{lattice}). 
Analogously to Fitting's bilattices, 
we show that trilattices associated with confidence levels are interlaced, 
making them interesting in their own right, from an algebraic point of view. 
In addition to 
providing an algebraic footing for our framework, trilattices also shed 
light on the relationship between our work and earlier works  
and offer useful insights. In particular, trilattices give rise to 
three ways of ordering confidence levels: the truth-order, where belief goes 
up and doubt comes down, the information order, where both belief and doubt 
go up, and the precision order, where the probability intervals associated 
with both belief and doubt become sharper, 
\ie the interval length decreases. 
This is to be contrasted with the known truth and information (called 
knowledge there) orders in a bilattice.

\item A purely lattice-theoretic basis for logic programming can 
be constructed using trilattices (similar to Fitting 
\cite{fitting-bilattice-lp-jlp91}). However, since our focus in this paper 
is probabilistic uncertainty,  we develop a  
probabilistic calculus for combining confidence levels 
associated with basic events into those for compound events based on them 
(Section \ref{prob-calc}).  Instead of committing to any specific rules 
for combining confidences, we propose a framework which allows a user to 
choose an appropriate ``mode" from a collection of available ones. 

\item We develop 
a generalized framework for rule-based programming with probabilistic 
knowledge, based on this calculus. We provide the declarative and fixpoint 
semantics for such programs and establish their equivalence (Section 
\ref{prob-ddb}). We also provide a sound and complete proof procedure 
(Section \ref{proof-theory}). 

\item We study the termination and complexity issues 
of such programs and show: (1) the closure ordinal of $T_P$ can be as high as 
$\omega$ in general (but no more), 
and (2) when only {\sl positive correlation} is used 
for disjunction\footnote{Other modes can be used (for conjunction/disjunction) 
in the ``non-recursive part" 
of the program.}, the data complexity of such programs is polynomial time. 
Our proof technique for the last result yields a similar result for van Emden's 
framework (Section \ref{termination}). 

\item We also compare our work with related work and bring 
out the advantages and generality of our approach (Section \ref{termination}). 
\end{itemize}


\section{Motivation} \label{motiv} 

In this section, we discuss the motivation for our work as well as comment on 
our design decisions for this framework. 
The motivation for using probability theory as opposed to 
other formalisms for representing uncertainty has been discussed at length in 
the literature 
\cite{carnap-prob-logic-bk62,Ng:PLP}. 
Probability theory is perhaps the 
best understood and mathematically well-founded paradigm in which uncertainty 
can be modeled and reasoned about. 
Two possibilities for associating probabilities with facts and rules in a DDB 
are van Emden's style of associating confidences with 
rules as a whole \cite{vEm86:QuanDedFPTh}, or the annotation style of 
Kifer and Subrahmanian \cite{KiSub92:GAPs}.
The second approach is more powerful:
It is shown in
\cite{KiSub92:GAPs} that the second approach can simulate the first.
The first approach, on the other hand,
has the advantage of intuitive appeal,
as pointed out by Kifer and Subrahmanian \cite{KiSub92:GAPs}.
In this paper, 
we choose the first approach. 
A comparison between our approach and annotation-based approach
with respect to termination and complexity issues is given in
Section \ref{termination}.

A second issue is whether we should insist on precise probabilities
or allow intervals (or ranges). Firstly, probabilities derived 
from any sources may have tolerances associated with them. Even experts 
may feel more comfortable with specifying a range rather than a precise 
probability. Secondly, Fenstad \cite{Fe80:ProbFOL} has shown (also see 
\cite{Ng:PLP}) that when enough information is not available about the 
interaction between events, the probability of compound events cannot be 
determined precisely: one can only give (tight) bounds. Thus, we associate 
ranges of probabilities with 
facts and rules.

A last issue is the following. Suppose (uncertain) 
knowledge contributed by an expert corresponds to the formula $F$. 
In general, we cannot assume the 
expert's knowledge is perfect. This means he does not necessarily know 
{\em all} situations in which $F$ holds. Nor does he know {\em all} situations 
where $F$ fails to hold (\ie $\neg F$ holds). He models the 
proportion of the situations where he knows $F$ holds as his {\sl belief} in 
$F$ and the proportion of situations where he knows $\neg F$ holds as his 
{\em doubt}. There could be situations, unknown to our expert, where $F$ 
holds (or $\neg F$ holds). These unknown situations correspond to the gap in 
his knowledge. Thus, as far as he knows, $F$ is {\em unknown} or 
{\em undefined} in these remaining situations. 
These observations, originally made by Fitting 
\cite{Fi88:LPLat}, give rise to the following definition.

\begin{defn}
\label{defn:cf}
({\em Confidence Level})
Denote by ${\cal C}[0, 1]$ the set of all closed subintervals over $[0, 1]$.
Consider the set $\conflat  =_{def} {\cal C}[0, 1] \times {\cal C}[0, 1]$.
A {\em Confidence Level} is an element of $\conflat$.
We denote a confidence level as $\conf$.
\end{defn}

In our approach confidence levels are associated with facts and rules.
The intended meaning of a fact (or rule) $F$ having a confidence $\conf$
is that $\alpha$ and $\beta$ are the lower and upper bounds of the
expert's {\em belief} in $F$, and
$\gamma$ and $\delta$ are the lower and upper bounds of the expert's
{\em doubt} in $F$.
These notions will be formalized in Section \ref{prob-calc}.

The following example illustrates such a scenario. (The figures in all 
our examples are fictitious.) 

\begin{ex}
Consider the results 
of Gallup polls conducted before the recent Canadian federal  
elections.

\noindent
1. Of the people surveyed, between 50\% and 53\% of the people
in the age group 19 to 30 favor the liberals. \\
2. Between 30\% and 33\% of the people in the above age group favor the 
reformists. \\
3. Between 5\% and 8\% of the above age group favor the tories. 

The reason we have ranges for each 
category is that usually some tolerance is associated with the results  
coming from such polls. 
Also, we do not make the proportion of undecided people 
explicit as our interest is in determining the support for the different 
parties. Suppose we assimilate the information above in a probabilistic 
framework. For each party, we compute the probability that a {\em randomly} 
chosen person from the sample population of the given age group will 
(not) vote for that party. 
We transfer this probability as the {\em subjective} probability that 
{\em any} person 
from that age group (in the actual population) will (not) vote for the party. 
The conclusions are given below, where $vote(X,P)$ says $X$ will vote for 
party $P$, $age$-$group1(X)$ says $X$ belongs to the age group specified above. 
$liberals$, $reform$, and $tories$ are constants, with the obvious 
meaning.

\noindent
1. $vote(X, liberals)$ $\prulearg{\langle[0.5, 0.53], [0.35, 0.41]\rangle}$
$age$-$group1(X)$. \\
2. $vote(X, reform)$: $\prulearg{\langle[0.3, 0.33], [0.55, 0.61]\rangle}$
$age$-$group1(X)$. \\
3. $vote(X, tories)$: $\prulearg{\langle[0.05, 0.08], [0.8, 0.86]\rangle}$
$age$-$group1(X)$. 

As usual, each rule is implicitly universally quantified outside the entire 
rule. 
Each rule is expressed in the form 
\[
A~\prule Body
\]
where $\alpha, 
\beta, \gamma, \delta \in [0,1]$. We usually require that $\alpha \leq \beta$ 
and $\gamma \leq \delta$. With each rule, we have associated two intervals. 
$[\alpha,\beta]$ ($[\gamma,\delta]$) is the {\em belief} ({\em doubt}) 
the expert has in the rule. 
Notice that from his knowledge, the expert can only conclude that 
the proportion of people he {\em knows} favor $reform$ or $tories$ will not 
vote for $liberals$. Thus the probability that 
a person in the age group 19-30 will not vote for liberals, according to the 
expert's knowledge, is in the range $[0.35, 0.41]$,   
obtained by summing the endpoints of the belief ranges for reform and 
tories. Notice that in this case $\alpha + \delta$ (or $\beta + \gamma$) 
is not necessarily $1$. This shows we cannot regard the expert's doubt as the  
complement (with respect to 1) of his belief. Thus, if we have to model what 
{\em necessarily} follows according to the expert's knowledge, 
then we must carry both 
the belief and the doubt explicitly. 
Note that this example
suggests just one possible means by which confidence levels
could be obtained from statistical data.
As discussed before, gaps in an expert's knowledge could
often directly result in both belief and doubt.
In general, there could be many ways in which both belief and doubt could
be obtained and associated with the basic facts. Given this, we believe that
an independent treatment of both belief and doubt is both necessary and
interesting for the purpose of obtaining the confidence levels associated
with derived facts. Our approach to independently capture belief and doubt
makes it possible to cope with incomplete knowledge regarding
the situations in which an event is true, false, or unknown
in a general setting.
\end{ex}

Kifer and Li \cite{Ki88:SemESUnc} 
and Baldwin \cite{baldwin-evidential-lp-jfuzzy87} have argued that 
incorporating  both belief and doubt (called disbelief there) is useful in 
dealing with incomplete knowledge, where different evidences may contradict 
each other. However, in their frameworks, doubt need not be maintained 
explicitly. For suppose we have a belief $b$ and a disbelief $d$ associated 
with a phenomenon. Then they can both be absorbed into one range $[b, 1-d]$ 
indicating that the effective certainty ranges over this set. 
The difference with our 
framework, however, is that we model what is {\em known} definitely, as 
opposed to what is {\em possible}. This makes (in our case) an explicit 
treatment of belief and doubt mandatory.


\section{The Algebra of Confidence Levels} \label{lattice}

Fitting \cite{fitting-bilattice-lp-jlp91} has shown that {\sl bilattices}  
(introduced by Ginsburg \cite{ginsburg-bilattice-compint88})  
lead to an elegant framework 
for quantified logic programming involving both belief and doubt. In this 
section, we shall see that a notion of {\sl trilattices} naturally arises 
with confidence levels. We shall establish the structure and properties of 
trilattices here, which will be used in later sections.

\begin{defn} \label{conf-lattice}
Denote by ${\cal C}[0, 1]$ the set of all closed subintervals over $[0, 1]$. 
Consider the set $\conflat  =_{def} {\cal C}[0, 1] \times {\cal C}[0, 1]$.  
We denote the elements of $\conflat$ as 
$\conf$. Define the following orders on this set. Let $\confi{1}$, $\confi{2}$ 
be any two elements of $\conflat$. 

\noindent
$\confi{1} \leq_t \confi{2}$~ iff ~ 
      $\alpha_1 \leq \alpha_2$, $\beta_1 \leq \beta_2$ ~and~ 
      $\gamma_2 \leq \gamma_1$, $\delta_2 \leq \delta_1$ \\
$\confi{1} \leq_k \confi{2}$ ~iff~ 
      $\alpha_1 \leq \alpha_2$, $\beta_1 \leq \beta_2$ ~and~ 
      $\gamma_1 \leq \gamma_2$, $\delta_1 \leq \delta_2$ \\
$\confi{1} \leq_p \confi{2}$ ~iff~
      $\alpha_1 \leq \alpha_2$, $\beta_2 \leq \beta_1$ ~and~ 
      $\gamma_1 \leq \gamma_2$, $\delta_2 \leq \delta_1$ 
\end{defn}

Some explanation is in order. The order $\leq_t$ can be considered the 
{\em truth} ordering: ``truth" relative to the expert's knowledge increases as 
belief goes up and doubt comes down. The order $\leq_k$ is the {\em knowledge} 
(or information) ordering: ``knowledge" (\ie the extent to which the expert 
commits his opinion on an assertion) increases as both belief and doubt 
increase. The order $\leq_p$ is the {\em precision} ordering: ``precision" of 
information supplied increases as the probability intervals become narrower. 
The first two orders are analogues of similar orders in bilattices. The third 
one, however, has no counterpart there.  
It is straightforward to see that each of the orders 
$\leq_t$, $\leq_k$, and $\leq_p$ is a partial order. 
$\conflat$ has a least and a greatest element with respect to each of these 
orders. In the following, we give the definition of meet and join 
with respect to the $\leq_t$ order. 
Operators with respect to the other orders have a similar definition. 

\begin{defn} 
\label{meet-join} 
Let $\langle\conflat, \leq_t, \leq_k, \leq_p\rangle$ be as defined in 
Definition \ref{conf-lattice}. 
Then the meet and join corresponding to the truth, knowledge (information), 
and precision orders are defined as follows. 
The symbols \meet\ and \join\ denote meet and join,
and the subscripts $t$, $k$, and $p$ represent
truth, knowledge, and precision, respectively.

\vs
\noindent
1. $\confi{1} \meett \confi{2}~=~$

\hfill $\langle[min\{\alpha_1, \alpha_2\}, min\{\beta_1, \beta_2\}],$ 
$[max\{\gamma_1, \gamma_2\}, max\{\delta_1, \delta_2\}]\rangle$.

\newpage
\noindent
2. $\confi{1} \joint \confi{2}~=~$

\hfill $\langle[max\{\alpha_1, \alpha_2\}, max\{\beta_1, \beta_2\}]$, 
$[min\{\gamma_1, \gamma_2\}, min\{\delta_1, \delta_2\}]\rangle$.

\noindent
3. $\confi{1} \meetk \confi{2}~=~$

\hfill $ \langle[min\{\alpha_1, \alpha_2\}, min\{\beta_1, \beta_2\}]$, 
$[min\{\gamma_1, \gamma_2\}, min\{\delta_1, \delta_2\}]\rangle$.

\noindent
4. $\confi{1} \joink \confi{2}~=~$

\hfill $ \langle[max\{\alpha_1, \alpha_2\}, max\{\beta_1, \beta_2\}]$,
$[max\{\gamma_1, \gamma_2\}, max\{\delta_1, \delta_2\}]\rangle$.

\noindent
5. $\confi{1} \meetp \confi{2}~=~$

\hfill $ \langle[min\{\alpha_1, \alpha_2\}, max\{\beta_1, \beta_2\}]$,
$[min\{\gamma_1, \gamma_2\}, max\{\delta_1, \delta_2\}]\rangle$.

\noindent
6. $\confi{1} \joinp \confi{2}~=~$

\hfill $ \langle[max\{\alpha_1, \alpha_2\}, min\{\beta_1, \beta_2\}]$,
$[max\{\gamma_1, \gamma_2\}, min\{\delta_1, \delta_2\}]\rangle$.

\vs
The top and bottom elements 
with respect to the various orders are as follows. 
The subscripts indicate the associated orders, as usual.

\vs
\noindent
$\top_t~=~\langle[1, 1], [0, 0]\rangle$, ~~~~~~~~$\bot_t~=~\langle[0, 0], [1, 1]\rangle$, \\
$\top_k~=~\langle[1, 1], [1, 1]\rangle$, ~~~~~~~~$\bot_k~=~\langle[0, 0], [0, 0]\rangle$,\\
$\top_p~=~\langle[1, 0], [1, 0]\rangle$, ~~~~~~~~$\bot_p~=~\langle[0, 1], [0, 1]\rangle$. 
\end{defn} 

\vs
$\top_t$ corresponds to total belief and no doubt; $\bot_t$ is the opposite. 
$\top_k$ represents maximal information (total belief and doubt), 
to the point of being probabilistically  inconsistent: belief and doubt 
probabilities sum to more than 1; $\bot_k$ gives the least information: no 
basis for belief or doubt; $\top_p$ is maximally precise, to the point of 
making the intervals empty (and hence inconsistent, in a non-probabilistic 
sense); $\bot_p$ is the least precise, as it imposes only trivial bounds 
on belief and doubt probabilities.

Fitting \cite{fitting-bilattice-lp-jlp91} defines a bilattice to be 
{\em interlaced} whenever the meet and join with respect to
any order of the bilattice 
are monotone with respect to the other order. 
He shows that it is the interlaced 
property of bilattices that makes them most useful and attractive. 
We say that a 
trilattice is {\em interlaced} provided the meet and join 
with respect to any order are 
monotone with respect to any other order. We have 

\begin{lemma} 
\label{lem:interlaced}
The trilattice $\langle\conflat, \leq_t, \leq_k, \leq_p\rangle$ defined above 
is interlaced. 
\end{lemma}

\noindent
{\bf Proof.} Follows directly from the fact that $max$ and $min$ are monotone 
functions. 
We show the proof for just one case. Let $\confi{1} \leq_p
\confi{3}$ and $\confi{2} \leq_p \confi{4}$. Then

\vs
\noindent
$\confi{1} \meet_t \confi{2} =$

\hfill $\langle[min\{\alpha_1, \alpha_2\},
min\{\beta_1, \beta_2\}], [max\{\gamma_1, \gamma_2\},
max\{\delta_1, \delta_2\}]\rangle $

\noindent
$\confi{3} \meet_t \confi{4} =$ 

\hfill $\langle[min\{\alpha_3, \alpha_4\},
min\{\beta_3, \beta_4\}], [max\{\gamma_3, \gamma_4\},
max\{\delta_3, \delta_4\}]\rangle$

\noindent
Since $\alpha_1 \leq \alpha_3,
\beta_3 \leq \beta_1, \alpha_2 \leq \alpha_4, \beta_4 \leq \beta_2$,
we have $min\{\alpha_1, \alpha_2\} \leq min\{\alpha_3, \alpha_4\}$, and
$min\{\beta_3, \beta_4\} \leq min\{\beta_1, \beta_2\}$. Similarly,
$min\{\gamma_1, \gamma_2\} \leq min\{\gamma_3, \gamma_4\}$ and
$min\{\delta_3, \delta_4\} \leq min\{\delta_1, \delta_2\}$. This implies

\negvs
\[\confi{1} \meett \confi{2} \leq_p \confi{3} \meett \confi{4}\]

\negvs
\noindent
Other cases are similar.
\qed

Trilattices are of independent interest in their own right, from an 
algebraic point of view. We also stress that they can be used as a basis 
for developing quantified/annotated logic programming schemes (which need 
not be probabilistic). This will be pursued in a future paper.

In closing this section, we note that other orders are also possible for
confidence levels. In fact, Fitting has shown that a fourth order,
denoted by $\leq_f$ in the following, 
together with the three orders defined above,
forms an interlaced ``quadri-lattice'' \cite{Fitting-letter}.
He also pointed out that this ``quadri-lattice'' can be generated
as the cross product of two bilattices.
Intuitively, a confidence level increases according to this fourth ordering,
when the   precision of the belief
component of a confidence level goes up, while that of the
doubt component goes down. That is,

\[
\confi{1} \leq_f \confi{2} \mbox{ iff }
\alpha_1 \leq \alpha_2, \beta_2 \leq \beta_1 \mbox{ and }
\gamma_2 \leq \gamma_1, \delta_1 \leq \delta_2
\]

In our opinion, the fourth order, while technically elegant, 
does not have the same intuitive appeal as the three orders -- 
truth, knowledge, and precision -- mentioned above. Hence, we 
do not consider it further in this paper. 
The algebraic properties of confidence levels
and their underlying lattices are interesting in their own right, and
might be used for developing alternative bases for quantitative logic
programming. This issue is orthogonal to the concerns of this paper.


\section{A Probabilistic Calculus} \label{prob-calc}

Given the confidence levels for (basic) 
events, how are we to derive the confidence levels for compound events 
which are based on them? Since we are working with probabilities,
our combination rules 
must respect probability theory.
We need a model of our knowledge 
about the interaction between events. A simplistic model studied in the 
literature (\eg see Barbara {\em et al.} \cite{Ba89:PRDM}) assumes 
{\em independence} between 
all pairs of events. This is highly restrictive and is of limited 
applicability. A general model, studied by Ng and Subrahmanian 
\cite{Ng:PLP,NgSub90:SFforSS+CP-in-DD} is 
that of {\em ignorance}: assume no knowledge about event interaction. Although 
this is the most general possible situation, it can be overly conservative  
when {\em some} knowledge is available, concerning some of the events. We 
argue that for ``real-life" applications, no single model of event interaction 
would suffice. Indeed, we need the ability to ``parameterize" the 
model used for event interaction, depending on what {\em is} known about the 
events themselves. In this section, we develop a probabilistic calculus which 
allows the user to select an appropriate ``mode" of event interaction, out 
of several choices, to suit his needs. 

Let {\bf L} be an arbitrary, but fixed, first-order language with 
finitely many constants, predicate symbols, infinitely many variables, 
and no function symbols \footnote{
In deductive databases, it is standard to restrict attention to
function free languages. Since input databases are finite
(as they are in reality), this leads to a finite Herbrand base.}.
We use (ground) atoms of {\bf L} to represent basic events. 
We blur the distinction between 
an event and the formula representing it. 
Our objective is to characterize confidence levels of boolean combinations 
of events involving the connectives $\neg, \wedge, \vee$, 
in terms of the confidence levels of the underlying basic events
under various modes (see below).

We gave an informal discussion of the meaning of confidence levels in
Section \ref{motiv}. We use the concept of {\em possible worlds}
to formalize the semantics of confidence levels.
 
\begin{defn}
\label{defn:semantics}
({\em Semantics of Confidence Levels})
According to the
expert's knowledge, an event $F$ can be true, false, or unknown.
This gives rise to 3 possible worlds.
Let $1, 0, \bot$ respectively denote
{\sl true}, {\sl false}, and {\sl unknown}.
Let $\worldi$ denote the world
where the truth-value of $F$ is $i$, $i \in \{0,1,\bot\}$,
and let $\wi$ denote the probability of the world $\worldi$
Then the assertion that the confidence level of $F$ is $\conf$,
written $\cf(F) = \conf$,
corresponds to the following constraints:
\begin{equation}
\label{eq:possibleWorlds}
\begin{array}{ccccl}
\alpha & \leq & \wo & \leq & \beta \\
\gamma & \leq & \wz & \leq & \delta\\
\wi & \geq & 0, & i & \in \{1,0,\bot\} \\
\Sigma_{i} \wi & = & 1 & &
\end{array}
\end{equation}
where $\alpha$ and $\beta$ are the lower and upper bounds of the
{\em belief} in $F$, and
$\gamma$ and $\delta$ are the lower and upper bounds of the
{\em doubt} in $F$.
\end{defn}

Equation (\ref{eq:possibleWorlds}) imposes certain restrictions on
confidence levels. 

\begin{defn}
\label{defn:consistentCF}
({\em  Consistent confidence levels})
We say a confidence level $\conf$ is {\em consistent}
if Equation (\ref{eq:possibleWorlds}) has an answer. 
\end{defn}

It is easily seen that:

\begin{prop}
\label{prop:consistentCF}
Confidence level $\conf$ is {\em consistent} provided
{\em (i)} $\alpha \leq \beta$ and $\gamma \leq \delta$, 
and {\em (ii)} $\alpha + \gamma \leq 1$.
\end{prop}

The consistency condition guarantees at least one solution to
Equation (\ref{eq:possibleWorlds}).
However, given a confidence level $\conf$,
there may be $\wo$ values in the $[\alpha,\beta]$
interval for which no $\wz$ value exists in the $[\gamma,\delta]$
interval to form an answer to Equation (\ref{eq:possibleWorlds}),
and vice versa.
We can ``trim'' the upperbounds of $\conf$ as follows to guarantee that
for each value in the $[\alpha,\beta]$ interval
there is at least one value in the $[\gamma,\delta]$ interval 
which form an answer to Equation (\ref{eq:possibleWorlds}).

\begin{defn}
\label{defn:reducedCF}
({\em Reduced confidence level})
We say a confidence level $\conf$ is {\em reduced} if 
for all $\wo \in [\alpha,\beta]$ there exist $\wz$, $\wb$ such that 
$\wo$, $\wz$, $\wb$ is a solution to Equation (\ref{eq:possibleWorlds}), and
for all $\wz \in [\gamma, \delta]$ there exist $\wo$, $\wb$ such that   
$\wo$, $\wz$, $\wb$ is a solution to Equation (\ref{eq:possibleWorlds}).
\end{defn}

It is obvious that a reduced confidence level is consistent.

\newpage
\begin{prop}
\label{prop:reducedCF}
Confidence level $\conf$ is {\em reduced} provided
{\em (i)} $\alpha \leq \beta$ and $\gamma \leq \delta$,
and {\em (ii)} 
$\alpha + \delta \leq 1$, and
$\beta + \gamma \leq 1$.
\end{prop}

\begin{prop}
\label{prop:reduction}
Let $c = \conf$ be a consistent confidence level.
Let $\beta' = 1 - \gamma$ and $\delta' = 1 - \alpha$.
Then, the confidence level
$c' = [\alpha,min(\beta,\beta')], [\gamma,min(\delta,\delta')]$
is a reduced confidence level.
Further, $c$ and $c'$ are probabilistically equivalent,
in the sense that they produce exactly the same answer sets 
to Equation (\ref{eq:possibleWorlds}).
\end{prop}

Data in a probabilistic deductive database, that is, facts and rules
that comprise the database, are associated with confidence levels.
At the atomic level, we require the confidence levels to be consistent.
This means each expert, or data source, should be consistent with
respect to the confidence levels it provides. This does not place
any restriction on data provided by different experts/sources,
as long as each is individually consistent. Data provided by different
experts/sources should be combined, using an appropriate combination mode
(discussed in next section). We will show that the combination
formulas for the various modes preserve consistent as well as
reduced confidence levels.

\subsection{Combination Modes}

Now, we introduce the various modes and characterize conjunction 
and disjunction under these modes. 
Let $F$ and $G$ represent two arbitrary 
ground (\ie variable-free) formulas. For a formula $F$, 
$\cf (F)$ will denote its confidence level.
In the following, we describe several interesting and natural
modes and establish some results on the confidence levels of 
conjunction and disjunction under these modes. 
Some of the modes are 
well known, although care needs to be taken to allow for 
the 3-valued nature of our framework. 

\vs
\noindent
{\sl 1. Ignorance:} This is the most general situation possible: nothing is 
assumed/known about event interaction between $F$ and $G$. 
The extent of the interaction between $F$ and $G$ could range from 
maximum overlap to minimum overlap. 

\vs
\noindent
{\sl 2. Independence:} This is a well-known mode. It simply says 
(non-)occurrence of one event does not influence that of the other.

\vs
\noindent
{\sl 3. Positive Correlation:} This mode corresponds to the knowledge that 
the occurrences of two events overlap as much as possible. This means the 
conditional probability of one of the events (the one with the larger 
probability) given the other is 1. 

\vs
\noindent
{\sl 4. Negative Correlation:} This is the exact opposite of positive 
correlation: the occurrences of the events overlap minimally. 

\vs
\noindent
{\sl 5. Mutual Exclusion:} This is a special case of negative correlation, 
where we know that the sum of probabilities of the events does not 
exceed 1. 

\vsp
We have the following results. 

\begin{prop} 
Let $F$ be any event, and let $\cf (F) = \conf$. Then $\cf (\neg F) =
\langle[\gamma, \delta],~[\alpha, \beta]\rangle$. Thus, negation
simply swaps belief and doubt.
\end{prop}

\noindent
{\bf Proof.}
Follows from the observation that $\cf (F) = \conf$ implies
that $\alpha \leq \wo \leq \beta$ and
$\gamma \leq \wz \leq \delta$,
where $\wo$ ($\wz$) denotes the probability of the
possible world where event $F$ is {\sl true} ({\sl false}).
\qed

\vs
The following theorem establishes the confidence levels of compound formulas 
as a function of those of the constituent formulas, under various modes. 

\begin{thm}
\label{thm:combinations}
Let $F$ and $G$ be any events and let 
$\cf (F) = \confi{1}$ and
$\cf (G) = \confi{2}$. Then the confidence
levels of the compound events $F\wedge G$ and $F \vee G$ are given as
follows. (In each case the subscript denotes the mode.)
 
\vs
\noindent
$\cf (F \wedge_{ig} G) =$

\hfill
$\langle[max\{0, \alpha_1 + \alpha_2 -1\}, min\{\beta_1, \beta_2\}],$
 $[max\{\gamma_1, \gamma_2\}$, $min\{1,$ $\delta_1 + \delta_2\}]\rangle$.

\noindent
$\cf (F \vee_{ig} G) =
\langle[max\{\alpha_1, \alpha_2\}, min\{1, \beta_1 + \beta_2\}],
 [max\{0, \gamma_1 + \gamma_2 -1\}$, $min\{\delta_1, \delta_2\}]\rangle$.

\noindent
$\cf (F \wedge_{ind} G) =
\langle[\alpha_1 \times \alpha_2, \beta_1 \times \beta_2],
[1 - (1 - \gamma_1) \times (1 - \gamma_2),
 1 - (1 - \delta_1) \times (1 - \delta_2)]\rangle$.

\noindent
$\cf (F \vee_{ind} G) =
\langle [1 - (1 - \alpha_1) \times (1 - \alpha_2),
  1 - (1 - \beta_1) \times (1 - \beta_2)],
[\gamma_1 \times \gamma_2, \delta_1 \times \delta_2]\rangle$.

\noindent
$\cf (F \wedge_{pc} G) =
\langle[min\{\alpha_1, \alpha_2\}, min\{\beta_1, \beta_2\}],
 [max\{\gamma_1, \gamma_2\}, max\{\delta_1, \delta_2\}]\rangle$.

\noindent
$\cf (F \vee_{pc} G) =
\langle[max\{\alpha_1, \alpha_2\}, max\{\beta_1, \beta_2\}],
 [min\{\gamma_1, \gamma_2\}, min\{\delta_1, \delta_2\}]\rangle$.

\noindent
$\cf (F \wedge_{nc} G) =$

\hfill
$\langle[max\{0, \alpha_1 + \alpha_2 - 1\}, max\{0, \beta_1 + \beta_2 - 1\}],
 [min\{1, \gamma_1 + \gamma_2\}, min\{1, \delta_1 + \delta_2\}]\rangle$.

\noindent
$\cf (F \vee_{nc} G) =$

\hfill
$\langle[min\{1, \alpha_1 + \alpha_2\}, min\{1, \beta_1 + \beta_2\}],
[max\{0, \gamma_1 + \gamma_2 - 1\}, max\{0, \delta_1 + \delta_2 - 1\}]\rangle$.

\noindent
$\cf (F \wedge_{me} G) =
\langle[0, 0], [min\{1, \gamma_1+\gamma_2\},
min\{1, \delta_1+\delta_2\}]\rangle$.

\noindent
$\cf (F \vee_{me} G) =
\langle[\alpha_1 + \alpha_2, \beta_1 + \beta_2],
 [max\{0, \gamma_1 + \gamma_2 - 1\},
max\{0, \delta_1 + \delta_2 - 1\}]\rangle$.
\end{thm} 

\noindent
{\bf Proof.} Each mode is characterized by a system of constraints, and
the confidence level of the formulas $F \wedge G, F \vee G$ are obtained by
extremizing certain objective functions subject to these constraints.
 
The scope
of the possible interaction between $F$ and $G$ can be characterized as
follows
(also see \cite{frechet-ignorance-prob-bounds-fund-math35,Ng:PLP}).
According to the
expert's knowledge, each of $F, G$ can be true, false, or unknown. This
gives rise to 9 possible worlds. Let $1, 0, \bot$ respectively denote
{\sl true}, {\sl false}, and {\sl unknown}. Let $\worldij$ denote the world
where the truth-value of $F$ is $i$ and that of $G$ is $j$, $i, j \in \{1, 0,
\bot\}$. 
\bigeg, $\worldoz$ is the world where $F$ is true and $G$ is false, while
$\worldzb$ is the world where $F$ is false and $G$ is unknown.
Suppose $\wij$
denotes the probability associated with world $\worldij$.
Then the possible scope
of interaction between $F$ and $G$ can be characterized by the following
constraints.
 
\begin{equation}
\begin{array}{lcccl}
\alpha_1 & \leq & \woz + \wob + \woo & \leq & \beta_1 \label{ign-eq} \\
\gamma_1 & \leq & \wzz + \wzb + \wzo & \leq & \delta_1 \\
\alpha_2 & \leq & \wzo + \wbo + \woo & \leq & \beta_2 \\
\gamma_2 & \leq & \wzz + \wbz + \woz & \leq & \delta_2 \\
\wij & \geq & 0, & i, j & \in \{1, 0,\bot\} \\
\Sigma_{i,j} \wij & = & 1 & &
\end{array}
\end{equation}
 
The above system of constraints must be satisfied for all modes. 
Specific constraints for various modes are obtained by adding
more constraints to those in 
Equation (\ref{ign-eq}). In all cases, the confidence
levels for $F \wedge G$ and $F\vee G$ are obtained as follows.
\begin{eqnarray*}
\cf (F\circ G) & = & \langle[min(\Sigma_{\worldij\models F\circ G} \wij),
                      max(\Sigma_{\worldij\models F\circ G} \wij)],\\
               &   & [min(\Sigma_{\worldij\not \models F\circ G} \wij),
                      max(\Sigma_{\worldij\not \models F\circ G} \wij)]\rangle
\end{eqnarray*}
where $\circ$ is $\wedge$ or $\vee$.

\vsp
\noindent
\underline{Case 1}: {\sl Ignorance}.
 
The constraints for ignorance are exactly those in 
Equation (\ref{ign-eq}).
The solution to the above linear program can be shown to be

\noindent
$\cf (F\wedge G) =
\langle[max\{0, \alpha_1 + \alpha_2 -1\}, min\{\beta_1, \beta_2\}],$
 $[max\{\gamma_1, \gamma_2\}$, $min\{1,$ $\delta_1 + \delta_2\}]\rangle$, 

\noindent
$\cf (F\vee G) =
\langle[max\{\alpha_1, \alpha_2\}, min\{1, \beta_1 + \beta_2\}],
 [max\{0, \gamma_1 + \gamma_2 -1\}$, $min\{\delta_1, \delta_2\}]\rangle$.

\noindent
The proof is very similar to the proof of a similar result in the context
of belief intervals (no doubt) by Ng and Subrahmanian \cite{Ng:PLP}.

\vsp
\noindent
\underline{Case 2}: {\sl Independence}.

Independence of events $F$ and $G$ can be characterized by
the equation $P(F | G) = P(F)$,
where $P(F | G)$ is the conditional probability
of the event $F$ given event $G$.
More specifically, since in our model
an event can be {\sl true}, {\sl false}, or {\sl unknown},
(in other words, we model belief and doubt independently)
we have \footnote{We have nine equations, but it can be shown that the other
five are dependent on these four}:
\begin{equation}
\begin{array}{lcccc}
P(F \mbox{ is true} & | & G \mbox{ is true}) & = &
P( F \mbox{ is true}) \\
P(F \mbox{ is true} & | & G \mbox{ is false}) & = &
P( F \mbox{ is true}) \\
P(F \mbox{ is false} & | & G \mbox{ is true}) & = &
P( F \mbox{ is false}) \\
P(F \mbox{ is false} & | & G \mbox{ is false}) & = &
P( F \mbox{ is false})
\end{array}
\end{equation}
Then the constraints characterizing independence is obtained
by adding the following equations to the system of constraints
(\ref{ign-eq}).
\begin{equation}
\begin{array}{lcc}
\label{ind-eq}
\woo & = & (\woz + \wob + \woo) \times (\wzo + \wbo + \woo) \\
\woz & = & (\woz + \wob + \woo) \times (\wzz + \wbz + \woz) \\
\wzo & = & (\wzz + \wzb + \wzo) \times (\wzo + \wbo + \woo) \\
\wzz & = & (\wzz + \wzb + \wzo) \times (\wzz + \wbz + \woz)
\end{array}
\end{equation}
The belief in $F \wedge_{ind} G$, and doubt in $F \vee_{ind} G$
can be easily verified from the system of constraints \ref{ign-eq}
and \ref{ind-eq}
\footnote{such {\em duality} results exist
with respect to conjunction and disjunctions lower and upper bounds
(and vice versa).}
\begin{equation}
\begin{array}{lcccc}
\alpha_1 \times \alpha_2 & \leq & \woo & \leq & \beta_1 \times \beta_2 \\
\gamma_1 \times \gamma_2 & \leq & \wzz & \leq & \delta_1 \times \delta_2
\end{array}
\end{equation}
To obtain the doubt in $F \wedge_{ind} G$ we need to compute
the minimum and maximum of
$\wzz + \wzb + \wzo + \woz + \wbz$.
It is easy to verify that:
\[
\gamma_1 + \gamma_2 - \gamma_1 \times \gamma_2 \leq
\wzz + \wzb + \wzo + \woz + \wbz \leq
\delta_1 + \delta_2 - \delta_1 \times \delta_2
\]
The belief in $F \vee_{ind} G$ is obtained similarly (in the dual manner.)
Thus, we have verified that

\noindent
$\cf (F \wedge_{ind} G) =
\langle[\alpha_1 \times \alpha_2, \beta_1 \times \beta_2],
[1 - (1 - \gamma_1) \times (1 - \gamma_2),
 1 - (1 - \delta_1) \times (1 - \delta_2)]\rangle$. \\
$\cf (F \vee_{ind} G) =
\langle [1 - (1 - \alpha_1) \times (1 - \alpha_2),
  1 - (1 - \beta_1) \times (1 - \beta_2)],
[\gamma_1 \times \gamma_2, \delta_1 \times \delta_2]\rangle$.

\vsp
\noindent
\underline{Case 3}: {\sl Positive Correlation}:
 
Two events $F$ and $G$ are positively correlated if
they overlap as much as possible.
This happens when
either ({\em i})
occurrence of $F$ implies occurrence of $G$,
or ({\em ii})
occurrence of $G$ implies occurrence of $F$.
In our framework we model belief and doubt independently,
and positive correlation is characterized by 4 possibilities:

\vs
\noindent
(a)
Occurrence of $F$ implies occurrence of $G$, and
non-occurrence of $G$ implies non-occurrence of $F$.
\\
(b)
Occurrence of $F$ implies occurrence of $G$, and
non-occurrence of $F$ implies non-occurrence of $G$.
\\
(c)
Occurrence of $G$ implies occurrence of $F$, and
non-occurrence of $G$ implies non-occurrence of $F$.
\\
(d)
Occurrence of $G$ implies occurrence of $F$, and
non-occurrence of $F$ implies non-occurrence of $G$.

\vs
Each of these four condition sets generates its own equations.
For example, (a)
can be captured by adding the following equations to the
system of constraints
\ref{ign-eq}.
 
\begin{equation}
\begin{array}{lcr}
\wob & = & 0  \\
\woz & = & 0  \\
\wbz & = & 0
\end{array}
\end{equation}
 
Hence, for condition (a), the system of constraints \ref{ign-eq} becomes
 
\begin{equation}
\begin{array}{lcccl}
\alpha_1 & \leq & \woo & \leq & \beta_1 \label{pc-eq} \\
\gamma_1 & \leq & \wzz + \wzb + \wzo & \leq & \delta_1 \\
\alpha_2 & \leq & \wzo + \wbo + \woo & \leq & \beta_2 \\
\gamma_2 & \leq & \wzz & \leq & \delta_2 \\
\wij & \geq & 0, & i, j & \in \{1, 0,\bot\} \\
\Sigma_{i,j} \wij & = & 1 & &
\end{array}
\end{equation}
 
The analysis is further complicated by the fact that the confidence levels
of $F$ and $G$ determine which of these cases apply, and it may
be different for the lowerbound and upperbound probabilities.
For example, if $\alpha_1 > \alpha_2$ ($\beta_1 > \beta_2$),
then the lowerbound (upperbound) for belief in $F \wedge_{pc} G$
is obtained when occurrence of $F$ implies occurrence of $G$.
Otherwise, these bounds are obtained when 
occurrence of $G$ implies occurrence of $F$.

The solution to these linear programs can be shown to be

\vs
\noindent
\hsp $\cf (F\wedge_{pc} G) =
\langle[min\{\alpha_1, \alpha_2\}, min\{\beta_1, \beta_2\}],
 [max\{\gamma_1, \gamma_2\}, max\{\delta_1, \delta_2\}]\rangle$,
and \\
\hsp $\cf (F \vee_{pc} G) =
\langle[max\{\alpha_1, \alpha_2\}, max\{\beta_1, \beta_2\}],
 [min\{\gamma_1, \gamma_2\}, min\{\delta_1, \delta_2\}]\rangle$.

\vs
A more intuitive approach to the derivation of confidence levels 
for conjunction and disjunction of positively correlated events
is to rely on the observation that these events overlap to the 
maximum extent possible. In our framework it means the worlds
where $F$ is {\em true} and those where $G$ is {\em true} overlap maximally,
and hence, one is included in the other.
Similarly, since we model belief and doubt independently,
the worlds where $F$ is {\em false} and those where $G$ is {\em false} also
overlap maximally. 
The combination formulas can be derived directly using these observations.

\vsp
\noindent
\underline{Case 4}: {\sl Negative Correlation}:

Negative correlation is an appropriate mode
to use whenever we know that events $F$ and $G$
overlap as little as possible.
This is to be contrasted with positive correlation,
where the extent of overlap is the greatest possible.
Mutual exclusion,
is a special case of negative correlation where
the sum of the probabilities of the two events
does not exceed 1. In this case the two events
do not overlap at all.
 
In the classical framework,
mutual exclusion of two events $F$ and $G$ is characterized 
by the statement: 
({\em i}) occurrence of $F$ implies non-occurrence of $G$,
and vice versa.
On the other hand, if the two events $F$ and $G$
are negatively correlated but not mutually exclusive, 
we have: 
({\em ii}) non-occurrence of $F$ implies occurrence of $G$,
and vice versa.
In Case ({\em i}) the sum of the probabilities
of the two events is at most 1, while
in Case ({\em ii}) this sum exceeds 1 and hence the two events
cannot be mutually exclusive.
In our framework we model belief and doubt independently,
and each of the above conditions translates to two conditions as follows.
Note that in our framework,  
``not {\em true}'' means ``{\em false} or {\em unknown}'',
and ``not {\em false}'' means ``{\em true} or {\em unknown}''. 

\noindent
\underline{Case (i):}  
 
\begin{itemize} 
\item[({\em a})]
Event $F$ is {\em true} implies $G$ is not {\em true}, and vice versa.
This condition generates the equation $\woo=0$.

\item[({\em b})]
The dual of condition (a),
when the non-occurrence of the two events don't overlap.
Event $F$ is {\em false} implies $G$ is not {\em false}, and vice versa.
This condition generates the equation $\wzz=0$.
\end{itemize} 

\noindent
\underline{Case (ii):}  

\begin{itemize} 
\item[({\em c})]
Event $F$ is not {\em true} implies $G$ is {\em true}, and vice versa.
This condition generates the equations
$\wzz = 0$, $\wbz = 0$, $\wzb = 0$, and $\wbb = 0$.

\item[({\em d})]
The dual of ({\em c}), 
$F$ is not {\em false} implies $G$ is {\em false}, and vice versa,
which generates the equations
$\woo = 0$, $\wbo = 0$, $\wob = 0$, and $\wbb = 0$.

\end{itemize}

Similar to the case for positive correlation,
the confidence levels of $F$ and $G$
determine which of these cases apply. For example, if
$\alpha_1+\alpha_2 > 1$, then case (c) should be used
to determine the lowerbound for belief in $F \wedge_{nc} G$. 

Alternatively, and more intuitively,
we can characterize negative correlation by observing that the worlds
where $F$ is {\em true} and those where $G$ is {\em true} overlap minimally,
and 
the worlds where $F$ is {\em false} and those where $G$ is {\em false} 
also overlap minimally.
The confidences of $F \wedge G$ and $F \vee G$ can be obtained using
the equations, or directly from the alternative characterization:

\noindent
$\cf (F \wedge_{nc} G) =$

\hfill
$\langle[max\{0, \alpha_1 + \alpha_2 - 1\}, max\{0, \beta_1 + \beta_2 - 1\}],
 [min\{1, \gamma_1 + \gamma_2\}, min\{1, \delta_1 + \delta_2\}]\rangle$

\noindent
$\cf (F \vee_{nc} G) =$

\hfill
$\langle[min\{1, \alpha_1 + \alpha_2\}, min\{1, \beta_1 + \beta_2\}],
[max\{0, \gamma_1 + \gamma_2 - 1\}, max\{0, \delta_1 + \delta_2 - 1\}]\rangle$

\vsp
\noindent
\underline{Case 5}: {\sl Mutual Exclusion}:

Mutual exclusion is a special case of
negative correlation. 
The main difference is that it requires the sum of the
two probabilities to be at most 1, which is not necessarily the case for
negative correlation (see the previous case).
In the classical framework, if two events are mutually exclusive,
their negation are not necessarily mutually exclusive.
Rather, they are negatively correlated. In our framework,
however, one or both conditions ({\em a}) and ({\em b}),
discussed in the previous case, can hold.
The appropriate condition is determined by the confidence levels of
the two mutually exclusive events, and the corresponding combination
formula can be obtained from the
combination formulas of negative correlation.
The following formulas, for example, are for
mutually exclusive events $F$ and $G$ where
$\alpha_1 + \alpha_2 \le \beta_1 + \beta_2 \le 1$
(but no other restriction).

\noindent
\hsp $\cf (F \wedge_{me} G) =
\langle[0, 0], [min\{1, \gamma_1+\gamma_2\},
min\{1, \delta_1+\delta_2\}]\rangle$. \\
\hsp $\cf (F \vee_{me} G) =
\langle[\alpha_1 + \alpha_2, \beta_1 + \beta_2],
 [max\{0, \gamma_1 + \gamma_2 - 1\},
max\{0, \delta_1 + \delta_2 - 1\}]\rangle$.

\qed

Next, we show that the combination formulas for various modes
preserve consistent as well as reduced confidence levels.
The case for reduced confidence levels is more involved and 
will be presented first.
The other case is similar, for which we only state the theorem.

\begin{thm} 
\label{thm:reduced}
Suppose $F$ and $G$ are any formulas and assume their confidence 
levels are reduced (Definition \ref{defn:reducedCF}).
Then the confidence levels of the formulas 
$F \wedge G$ and $F \vee G$, obtained under the various modes above 
are all reduced. 
\end{thm}

\noindent
{\bf Proof.}
Let $\cf (F) = \confi{1}$ and $\cf (G) = \confi{2}$.
Since the confidence levels of $F$ and $G$ are reduced, 
we have:

\noindent
\hsp $0 \leq \alpha_i \leq \beta_i \leq 1$  \\
\hsp $0 \leq \gamma_i \leq \delta_i \leq 1$  \\
\hsp $\alpha_i + \delta_i \leq 1$  \\
\hsp $\beta_i + \gamma_i \leq 1$\\
The consistency of the confidence levels of the
combination events $F \wedge G$ and $F \vee G$ in different modes as
derived in Theorem \ref{thm:combinations} follow from 
the above constraints. For example
let us consider

\[
\cf (A \wedge_{ig} B) =
\langle[max\{0, \alpha_1 + \alpha_2 -1\}, min\{\beta_1, \beta_2\}],
 [max\{\gamma_1, \gamma_2\}, min\{1, \delta_1 + \delta_2\}]\rangle
\]
We need to show

\noindent
\hsp (1) $max\{0, \alpha_1 + \alpha_2 -1\} \leq min\{\beta_1, \beta_2\}$\\
\hsp (2) $max\{\gamma_1, \gamma_2\} \leq min\{1, \delta_1 + \delta_2\}$\\
\hsp (3) $max\{0, \alpha_1 + \alpha_2 -1\} + 
min\{1, \delta_1 + \delta_2\} \leq 1$\\
\hsp (4) $min\{\beta_1, \beta_2\} + max\{\gamma_1, \gamma_2\} \leq 1$

\vs
To prove (1): If $max\{0, \alpha_1 + \alpha_2 -1\} = 0$ then (1) holds.
Otherwise, assume, without loss of generality, 
that $min\{\beta_1, \beta_2\} = \beta_1$. We can write

\noindent
\hsp $\alpha_1 \leq \beta_1$\\
\hsp $\alpha_2 \leq 1$

\noindent
and hence

\noindent
\hsp $\alpha_1 + \alpha_2 \leq \beta_1 +1$

\noindent
and (1) follows. 

\vs
\noindent
Inequality (2) follows easily from $\gamma_i \leq \delta_i \leq 1$.

\vs
\noindent
To prove (3): If $max\{0, \alpha_1 + \alpha_2 -1\} = 0$ then (3) holds.
Otherwise, we can write

\noindent
\hsp $\alpha_1 + \delta_1 \leq 1$  \\
\hsp $\alpha_2 + \delta_2 \leq 1$

\noindent
and hence

\noindent
\hsp $\alpha_1 + \alpha_2 -1 + \delta_1 + \delta_2 \leq 1$

\noindent
and (3) follows. Note that if $\delta_1 + \delta_2 > 1$ then
$\alpha_1 + \alpha_2 -1 + 1 \leq 1$ follows from the above constraint.

\vs
\noindent
To prove (4) let
$min\{\beta_1, \beta_2\} = \beta_j$
and
$max(\gamma_1, \gamma_2\} = \gamma_k$
where $j, k \in \{1, 2\}$.
Then
$\beta_j + \gamma_k \leq \beta_k + \gamma_k \leq 1$.

\vs
Proving the consistency of the confidence levels of other combinations
and other modes are similar and will not be elaborated here.
\qed

\begin{thm}
\label{thm:consistent}
Suppose $F$ and $G$ are any formulas and assume their confidence
levels are consistent (Definition \ref{defn:consistentCF}).
Then the confidence levels of the formulas
$F \wedge G$ and $F \vee G$, obtained under the various modes above
are all consistent.
\end{thm}

\noindent
{\bf Proof.}
Proof is similar to the previous theorem and is omitted.
\qed


\section{Probabilistic Deductive Databases} \label{prob-ddb}

In this section, we develop a framework for probabilistic deductive 
databases using a language of probabilistic programs (p-programs). 
We make use of the probabilistic calculus developed in 
Section \ref{prob-calc} and develop the syntax and declarative semantics for 
programming with confidence levels. We also provide the fixpoint 
semantics of programs in this framework and establish its equivalence to 
the declarative semantics. 
We will use the first-order language {\bf L} of Section \ref{prob-calc} as 
the underlying logical language in this section. 

\vsp
\noindent
{\bf Syntax of p-Programs:} A {\em rule} is an expression of the form 
$A~\shortprule~B_1, \ldots, B_m$, $m \geq 0$, 
where $A, B_i$ are atoms and $c \equiv \conf$ 
is the 
confidence level associated with the rule\footnote{We assume only 
consistent confidence levels henceforth (see Section \ref{prob-calc}).}.  
When $m = 0$, we call this a 
{\em fact}. All variables in the rule are assumed to be universally quantified 
outside the whole rule, as usual. We restrict attention to range restricted 
rules, as is customary. 
A {\em probabilistic rule} (p-rule) is a triple 
$(r;~~\mu_r,\mu_p)$, where $r$ is a rule, $\mu_r$ is a mode 
indicating how to conjoin the confidence levels of the subgoals in the body 
of $r$ (and with that of $r$ itself), and $\mu_p$ is a mode indicating how 
the confidence levels of different derivations of an atom involving the 
head predicate of $r$ are to be disjoined. We say $\mu_r$ ($\mu_p)$ is the 
mode associated with the body (head) of $r$,
and call it the {\em conjunctive} ({\em disjunctive}) mode. 
We refer to $r$ as the underlying rule of this p-rule. 
When $r$ is a fact, we omit $\mu_r$ for obvious reasons. 
A {\sl probabilistic program} (p-program) is a 
finite collection of p-rules such that whenever there are p-rules whose 
underlying rules define the same predicate, the mode associated with their 
head is identical. 
This last condition ensures different rules defining the 
same predicate $q$ agree on the manner in which confidences of 
identical $q$-atoms generated by these rules are to be 
combined. 
The notions of Herbrand universe $H_P$ and Herbrand base $B_P$ associated 
with a p-program $P$ are defined as usual. A p-rule is ground exactly 
when every atom in it is ground. The Herbrand instantiation $P^*$ 
of a p-program is defined in the obvious manner.
The following example illustrates our framework. 

\begin{ex} \label{medical-ex} 
People are assessed to be at high risk for various diseases, depending on 
factors such as age group, family history (with respect to the disease), etc. 
Accordingly, high risk patients are administered appropriate medications, 
which are prescribed by doctors among several alternative ones.  
Medications cause side effects, sometimes harmful 
ones, leading to other symptoms and diseases\footnote{Recent studies on the 
effects of certain medications on high risk patients for breast cancer 
provide one example of this.}. 
Here, the extent of risk, administration of 
medications\footnote{Uncertainty in this is mainly caused by the choices 
available and the fact that even under identical conditions doctors need not 
prescribe the same drug. The probabilities here can be derived from 
statistical data on the  relative frequency of prescriptions of drugs under 
given conditions.}, side effects (caused by medications), and prognosis are all 
uncertain phenomena, and we associate confidence levels with them. 
The following program is a sample of the uncertain knowledge related to these 
phenomena.

\vs
\noindent
1. $(high$-$risk(X, D)$ 
           $\prulearg{\langle[0.65,0.65], [0.1,0.1]\rangle}$  
           $midaged(X)$, $family$-$history(X, D);$ $ind,\_)$. \\ 
2. $(takes(X, M)$ 
           $\prulearg{\langle[0.40,0.40], [0,0]\rangle}$
           $high$-$risk(X, D)$, $medication(D, M);$ $ign,\_)$.\\
3. $(prognosis(X, D)$ 
           $\prulearg{\langle[0.70,0.70], [0.12,0.12]\rangle}$
           $high$-$risk(X, D);ign,pc)$.\\
4. $(prognosis(X, D)$ 
           $\prulearg{\langle[0.20,0.20], [0.70,0.70]\rangle}$
           $takes(X, M)$, $side$-$effects(M, D);$ $ind,pc)$.

\vs
We can assume an appropriate set of facts (the EDB) in conjunction with the 
above program. For rule 1, it is easy to see that each ground atom involving 
the predicate $high$-$risk$ has at most one derivation. Thus, a disjunctive 
mode for this rule will be clearly redundant, and we have suppressed it for 
convenience. A similar remark holds for rule 2. Rule 1 says that if a person 
is midaged and the disease $D$ has struck his ancestors, then the confidence 
level in the person being at high risk for $D$ is given by propagating the 
confidence levels of the body subgoals and combining them with the rule 
confidence in the sense of $\wedge_{ind}$. This could be  based on an expert's 
belief that the factors $midaged$ and $family$-$history$ contributing to 
high risk for the disease are independent. Each of the other rules 
has a similar explanation. For the last rule, we note that the potential 
of a medication to cause side effects is an intrinsic property independent of 
whether one takes the medication. Thus the conjunctive mode used there is 
independence. 
Finally, note that rules 3 and 4, defining $prognosis$,
use positive correlation as a conservative way of combining confidences
obtained from different derivations.
For simplicity, we show each interval in the above rules 
as a point probability. Still, note that the confidences for atoms derived from the program will be genuine intervals. 
\qed
\end{ex} 

\noindent
{\bf A Valuation Based Semantics:} We develop the declarative semantics 
of p-programs based on the notion of valuations. Let $P$ be a p-program. 
A {\em probabilistic 
valuation} is a function $v:~B_P \ra \conflat$ which associates a 
confidence level with each ground atom in $B_P$. We define the satisfaction 
of p-programs under valuations, with respect to the truth order $\leq_t$ of the 
trilattice (see Section \ref{prob-calc})\footnote{Satisfaction can be defined 
with respect to each of the 3 orders of the trilattice, 
giving rise to different 
interesting semantics. Their discussion is beyond the scope of this paper.}. 
We say a valuation $v$ {\em satisfies} a ground p-rule 
$\rho \equiv (\shortstdrule;~\mu_r,\mu_p)$, denoted $\models_v \rho$, 
provided $c \andmode v(B_1) \andmode 
\cdots \andmode$ $v(B_m) \leq_t v(A)$. The intended meaning is that 
in order to satisfy this p-rule, $v$ must assign a confidence level to 
$A$ that is no less true (in the sense of $\leq_t$) than the result 
of the conjunction of the confidences assigned to $B_i$'s by $v$ and the rule 
confidence $c$, in the sense of the mode $\mu_r$. 
Even when a 
valuation satisfies (all ground instances of) each rule in a p-program, it 
may not satisfy the p-program as a whole. The reason is that confidences 
coming from different derivations of atoms are combined strengthening the 
overall confidence. Thus, we need to impose the following additional 
requirement. 

Let $\rho \equiv (r \equiv \shortstdrule;~~\mu_r,\mu_p)$ be a ground 
p-rule, and $v$ a valuation. Then we denote by $rule$-$conf(A,\rho,v)$ 
the confidence level propagated to the head of this rule under the 
valuation $v$ and the rule mode $\mu_r$, given by the expression 
$c \andmode v(B_1) \andmode \cdots \andmode v(B_m)$.  
Let $P^* = P_1^* \cup \cdots \cup P_k^*$ be the partition of $P^*$ such that 
(i) each $P_i^*$ contains all (ground) p-rules which define the same atom, 
say $A_i$, and (ii) $A_i$ and $A_j$ are distinct, whenever $i \neq j$. Suppose 
$\mu_i$ is the mode associated with the head of the p-rules in $P_i^*$. 
We denote by $atom$-$conf(A_i,P,v)$ the confidence level determined for 
the atom $A_i$ under the valuation $v$ using the program $P$. This is 
given by the expression $\vee_{\mu_i} \{rule$-$conf(A_i,\rho,v) | \rho \in 
P_i*\}$. We now define satisfaction of p-programs. 

\begin{defn}
\label{def:satisfaction}
Let $P$ be a p-program and $v$ a valuation. Then $v$ satisfies $P$, denoted 
$\models_v P$ exactly when $v$ satisfies each (ground) p-rule in $P^*$, and 
for all atoms $A \in B_P$, $atom$-$conf(A,P,v) \leq_t v(A)$. 
\end{defn}

The additional requirement ensures the valuation assigns a strong enough 
confidence to each atom so it will support the combination of 
confidences coming from a number of rules (pertaining to this atom). A 
p-program $P$ logically implies a p-fact $\shortstdfact$, denoted 
$P \models \shortstdfact$, provided every valuation satisfying $P$ also 
satisfies $\shortstdfact$. We next have 

\begin{prop}
Let $v$ be a valuation and $P$ a p-program. Suppose the mode associated 
with the head of each p-rule in $P$ is positive correlation. Then 
$\models_v P$ iff $v$ satisfies each rule in $P^*$. 
\end{prop}

\noindent
{\bf Proof}.
We shall show that if $rule$-$conf(A,\rho_i,v) \leq_t v(A)$ for all rules
$\rho_i$ defining a ground atom $A$, then
$atom$-$conf(A,P,v) \leq_t v(A)$,
where the disjunctive mode for $A$ is positive correlation.
This follows from the formula for $\vee_{pc}$, 
obtained in Theorem \ref{thm:combinations}.
It is easy to see that $c_1 \vee_{pc} c_2 = c_1 \joint c_2$.
But then, $rule$-$conf(A,\rho_i,v) \leq_t v(A)$ implies that
$\joint \{ rule$-$conf(A,\rho_i,v) \} \leq_t v(A)$
and hence $atom$-$conf(A,P,v) \leq_t v(A)$.
\qed

\vs
The above proposition shows that when positive correlation is the only 
disjunctive mode used, satisfaction is very similar to the classical 
case. 

\vs
For the declarative semantics of p-programs, we need something like the 
``least" valuation satisfying the program. It is straightforward to 
show that the class of all valuations $\Upsilon$ from $B_P$ to $\conflat$ 
itself forms a trilattice, complete with all the 3 orders  and the 
associated meets and joins. They are obtained by a pointwise extension of 
the corresponding order/operation on the trilattice $\conflat$. We give 
one example. For valuations $u, v$, $u \leq_t v$ iff $\forall A \in B_P$, 
$u(A) \leq_t v(A)$; $\forall A \in B_P$, $(u \meett v)(A) = u(A) \meett v(A)$. 
One could investigate ``least" 
with respect to each of the 3 orders of the trilattice. In this paper, we 
confine attention to the order $\leq_t$. The least (greatest) valuation is then 
the valuation {\bf false} ({\bf true}) which assigns the confidence 
level $\bot_t$ ($\top_t$) to every ground atom. We now have 

\begin{lemma}
\label{lem:lvaluation}
Let $P$ be any p-program and $u, v$ be any valuations satisfying $P$. Then 
$u \meett v$ is also a valuation satisfying $P$. In particular, 
$\meett \{v \mid \  \models_v P\}$ is the least valuation satisfying $P$. 
\end{lemma}

\noindent
{\bf Proof.}
We prove this in two steps. First, we show that for any ground p-rule \\
\hsp $\rho \equiv (r \equiv \shortstdrule;~~\mu_r,\mu_p)$ \\
whenever valuations 
$u$ and $v$ satisfy $\rho$, so does $u \meett v$. Secondly, we shall show 
that for a p-program $P$, whenever {\it atom-conf\/}$(A, P, u) \leq_t u(A)$ 
and  {\it atom-conf\/}$(A, P, v) \leq_t v(A)$,
then we also have 
{\it atom-conf\/}$(A, P, u\meett v) \leq_t u\meett v(A)$. 
The lemma will follow 
from this. 

\vs
\noindent
{\sf (1)} Suppose $u \models \rho$ and $v \models \rho$. We prove the  
case where the conjunctive mode $\mu_r$ associated with this rule is 
ignorance. The other cases are similar. It is straightforward 
to verify the following. 

\noindent
\hsp 
(i) $u\meett v (B_1) \wedge_{ig} \cdots \wedge_{ig} u\meett v (B_m) \leq_t 
u(B_1) \wedge_{ig} \cdots \wedge_{ig} u(B_m) \leq_t u(A)$. \\ 
\hsp 
(ii) $u\meett v (B_1) \wedge_{ig} \cdots \wedge_{ig} u\meett v (B_m) \leq_t 
v(B_1) \wedge_{ig} \cdots \wedge_{ig} v(B_m) \leq_t v(A)$.

\noindent
From (i) and (ii), we have 
$u\meett v (B_1) \wedge_{ig} \cdots \wedge_{ig} u\meett v (B_m) \leq_t 
u\meett v(A)$, showing $u\meett v \models \rho$. 

\vs
\noindent
{\sf (2)} Suppose $u, v$ are any two valuations satisfying a p-program $P$. 
Let $\rho_1, \ldots, \rho_n$ be the set of all ground p-rules 
in $P^*$ whose heads are $A$. Let $c_i = $ {\it rule-conf\/}$(A, \rho_i, u)$ 
and $d_i = $ {\it rule-conf\/}$(A, \rho_i, v)$. Since $u \models P$ and 
$v \models P$,
we have that $\vee_{\mu_p} (c_i \mid 1\leq i\leq n) \leq_t u(A)$ and 
$\vee_{\mu_p} (d_i \mid 1\leq i\leq n) \leq_t v(A)$, where $\mu_p$ is the 
disjunctive mode associated with $A$.
Again, we give the proof for the case 
$\mu_p$ is ignorance as the other cases are similar. Let $e_i = $
{\it rule-conf\/}$(A, \rho_i, u\meett v), 1\leq i\leq n$. Clearly, 
$e_i \leq_t c_i$ and $e_i \leq_t d_i$.
Thus, $\vee_{\mu_p} (e_i \mid 1\leq i\leq n) 
\leq_t \vee_{\mu_p} (c_i \mid 1\leq i\leq n) \leq_t u(A)$ and 
$\vee_{\mu_p} (e_i \mid 1\leq i\leq n)
\leq_t \vee_{\mu_p} (d_i \mid 1\leq i\leq n) \leq_t 
v(A)$. It then follows that 
$\vee_{\mu_p} (e_i \mid 1\leq i\leq n) \leq_t 
u(A) \meett v(A) = u\meett v(A)$, which was to be shown. \qed 
 
\vs
We take the least valuation satisfying a p-program as characterizing its 
declarative semantics. 

\begin{ex}
Consider the following p-program $P$. 

\noindent
1. $(A~\prulearg{\langle[0.5,0.7],~[0.3,0.45]\rangle}~B;~~ind,~pc)$. 
\hfill
2. $(A~\prulearg{\langle[0.6,0.8],~[0.1,0.2]\rangle}~C;~~ign,~pc)$. \\
3. $(B~\prulearg{\langle[0.9,0.95],~[0,0.1]\rangle};~~\_,~ind)$. ~~~~~~
4. $(C~\prulearg{\langle[0.7,0.8],~[0.1,0.2]\rangle};~~\_,~ind)$. 

\vs
In the following we show three valuations $v_1, v_2, v_3$, of which $v_1$ 
and $v_3$ satisfy $P$, while $v_2$ does not. In fact, $v_3$ is the least 
valuation satisfying $P$.

\vs
$\begin{array}{cccc}
val & B & C & A \\
v_1 & \langle[0.9,1],~[0,0]\rangle & \langle[0.8,0.9],~[0.05,0.1]\rangle & 
                \langle[0.5,0.9],~[0,0]\rangle \\
v_2 & \langle[0.9,1],~[0,0]\rangle & \langle[0.9,1],~[0,0]\rangle & 
                \langle[0.5,0.7],~[0.1,0.4]\rangle \\ 
v_3 & \langle[0.9,0.95],~[0,0.1]\rangle & \langle[0.7,0.8],~[0.1,0.2]\rangle 
            &   \langle[0.45,0.8],~[0.1,0.4]\rangle
\end{array}$

\vs
For example, consider $v_1$.
It is easy to verify that $v_1$ satisfies $P$.
Rules 1 through 4 are satisfied by $v_1$ since:

\vs
\noindent
$\langle[0.5,0.7],~[0.3,0.45]\rangle \wedge_{ind}
 \langle[0.9,1],~[0,0]\rangle =
 \langle[0.45,0.7],~[0,0]\rangle \le_t
 \langle[0.5,0.9],~[0,0]\rangle$

\noindent
$\langle[0.6,0.8],~[0.1,0.2]\rangle \wedge_{ign}
 \langle[0.8,0.9],~[0.05,0.1]\rangle =$

\hfill
$\langle[0.4,0.8],~[0.1,0.3]\rangle \le_t
 \langle[0.5,0.9],~[0,0]\rangle$

\noindent
$\langle[0.9,0.95],~[0,0.1]\rangle \le_t \langle[0.9,1],~[0,0]\rangle$

\noindent
$\langle[0.7,0.8],~[0.1,0.2]\rangle \le_t \langle[0.8,0.9],~[0.05,0.1]\rangle$

\vs
Further, the confidence level of $A$ computed by the
combination of rules 1 and 2 is also satisfied by $v_1$,
namely,

\vs
\noindent
$( \langle[0.5,0.7],~[0.3,0.45]\rangle \wedge_{ind}
   \langle[0.9,1],~[0,0]\rangle )        \vee_{pc}
 ( \langle[0.6,0.8],~[0.1,0.2]\rangle$

\hfill
$ \wedge_{ign}
   \langle[0.8,0.9],~[0.05,0.1]\rangle) = 
   \langle[0.45,0.8],~[0,0] \le_t
   \langle[0.5,0.9],~[0,0]\rangle$ 
\qed
\end{ex}

\noindent
{\bf Fixpoint Semantics:} We associate an ``immediate consequence" 
operator $T_P$ with a p-program $P$, defined as follows.

\begin{defn} \label{tp-defn}
Let $P$ be a p-program and $P^*$ its Herbrand instantiation. 
Then $T_P$ is a function $T_P:~\Upsilon \ra \Upsilon$, defined as follows. 
For any probabilistic valuation $v$, and any ground atom $A \in B_P$, 
$T_P(v)(A) = \vee_{\mu_p} \{c_A | $ there exists a p-rule 
$(\shortstdrule,~\mu_r,~\mu_p)  \in P^*,$ such that
$c_A = c \andmode v(B_1) \andmode \cdots \andmode v(B_m)\}$. 
\end{defn}

Call a valuation $v$ {\em consistent} provided for every atom $A$, $v(A)$ is  
consistent, as defined in Section \ref{lattice}. 

\begin{thm}
\label{thm:tpmonotone}
(1) $T_P$ always maps consistent valuations to consistent valuations.  
(2) $T_P$ is monotone and continuous. 
\end{thm}

\noindent
{\bf Proof.}
(1) This fact follows Theorem \ref{thm:consistent}, where
we have shown that the combination functions for all modes
map consistent confidence levels to consistent confidence levels.
(2) This follows from the fact that the combination functions for all modes
are themselves monotone and continuous.
\qed

\vsp
We define bottom-up iterations based on $T_P$ in the usual manner. \\
$T_P\uparrow 0 = {\bf false}$ (which assigns the truth-value $\bot_t$ to 
every ground atom).    \\ 
$T_P\uparrow \alpha = T_P(T_P\uparrow \alpha-1)$, for a successor ordinal 
$\alpha$. \\
$T_P\uparrow \alpha = \joint \{T_P\uparrow \beta | \beta < \alpha\}$, for a 
limit ordinal $\alpha$. 

\vs
We have the following results. 

\begin{prop}
\label{prop:fixpoint}
Let $v$ be any valuation and $P$ be a p-program. Then $v$ satisfies $P$ 
iff $T_P(v) \leq_t v$. 
\end{prop}

\noindent
{\bf Proof.}
{\em (only if).}
If $v$ satisfies $P$, then by Definition \ref{def:satisfaction},
for all atoms $A \in B_P$, $atom$-$conf(A,P,v) \leq_t v(A)$
and hence $T_P(v) \leq_t v$.

\noindent
{\em (if).}
If $T_P(v) \leq_t v$, then by the definition of $T_P$
(Definition \ref{tp-defn})
for all atoms $A \in B_P$, $atom$-$conf(A,P,v) \leq_t v(A)$
and hence $v$ satisfies $P$.
\qed

\vsp
The following theorem is the analogue of the van Emden-Kowalski theorem for 
classical logic programming. 

\begin{thm} 
\label{thm:lfplvaluation}
Let $P$ be a p-program. Then the following claims hold. \\ 
(i) $lfp(T_P) = \meett\{v | \models_v P\} =$ the $\leq_t$-least valuation 
satisfying $P$. \\ 
(ii) For a ground atom $A$, $lfp(T_P)(A) = c$ iff $P \models \shortstdfact$. 
\end{thm}

\noindent
{\bf Proof.}
Follows Lemma \ref{lem:lvaluation},  Theorem \ref{thm:tpmonotone} 
and Proposition \ref{prop:fixpoint}.
Proof is similar to the analogous theorem of logic programming
and details are omitted.
\qed

\section{Proof Theory} \label{proof-theory} 

Since confidences coming from different derivations of a fact are combined, 
we need a notion of disjunctive proof-trees. We note that the 
notions of substitution, unification, etc. are analogous to the classical ones. A variable appearing in a rule is {\em local} if it only appears in its body. 

\newpage
\begin{defn} \label{dpt} 
Let $G$ be a(n atomic) goal and $P$ a p-program. Then a 
{\sl disjunctive proof-tree} 
(DPT) for $G$ with respect to $P$ is a tree $T$ defined as follows. \\
1. $T$ has two kinds of nodes: {\em rule} nodes and {\em goal} nodes. 
Each rule node is 
labeled by a rule in $P$ and a substitution. Each goal node is 
labeled by an atomic goal. The root is a goal node labeled $G$. \\
2. Let $u$ be a goal node labeled by an atom $A$. Then every child (if any) of 
$u$ is a rule node labeled $(r, \theta)$, 
where $r$ is a rule in $P$ whose head is unifiable with $A$ 
using the mgu $\theta$. 
We assume that each time a rule $r$ appears in the tree,
its variables are renamed to new variables
that do not appear anywhere else in the tree.
Hence $r$ in the label $(r, \theta)$ actually represents a
renamed instance of the rule. \\
3. If $u$ is a rule node labeled $(r, \theta)$, 
then whenever an atom $B$ occurs in the body of $r' = r \theta$,
$u$ has a goal child $v$ labeled $B$. \\
4. For any two substitutions $\pi, \theta$ occurring in $T$, $\pi(V) = 
\theta(V)$, for every variable $V$.
In other words, all substitutions occurring 
in $T$ are compatible.                      
\end{defn}

A node without children is called a leaf.
A {\sl proper} DPT is a finite DPT $T$ such that 
whenever $T$ has a goal leaf labeled $A$,
there is no rule in $P$ whose head is unifiable with $A$.
We only consider proper DPTs unless otherwise 
specified. A rule leaf is a {\sl success} node
(represents a database fact)
while a goal leaf is a {\sl failure} node.

\vs
\noindent
{\bf Remarks:} 

\noindent
(1) The definition of disjunctive proof tree captures the idea
that when working with uncertain information in the form of probabilistic
rules and facts, we need to consider the disjunction of all proofs
in order to determine the best possible confidence in the goal being
proved.

\noindent
(2) However, notice that the definition does {\em not} insist that a goal node 
$A$ should have rule children corresponding to all possible unifiable 
rules and mgu's. 

\noindent
(3) We assume without 
loss of generality that all rules in the p-program are standardized apart by 
variable renaming so they share no common variables. 

\noindent
(4) A goal node can have several rule children corresponding to
the same rule. That is, a goal node can have children labeled
$(r, \theta_1), \ldots, (r, \theta_n)$,
where $r$ is (a renamed version of) a rule in the program.
But we require that $r'_i = r \theta_i$, $i = 1, \ldots, n$,
be distinct.

\noindent
(5) 
We require all substitutions in the tree to be compatible. The convention 
explained above ensures there will be no conflict among them on account of 
common variables across rules (or different invocations of the same rule). 

\noindent
(6) Note that a DPT can be finite or infinite. 

\noindent
(7) In a proper DPT, goal leaves are necessarily failure 
nodes; this is not true in non-proper DPTs.

\noindent
(8) A proper DPT with no failure nodes has only rule leaves,
hence, it has an odd height.

\vsp
Confidences are associated with (finite) DPTs as follows. 

\begin{defn} \label{dpt-conf} 
Let $P$ be a p-program, $G$ a goal, and $T$ any finite DPT for $G$ 
with respect to $P$. 
We associate confidences with the nodes of $T$ as follows. \\
1. Each failure node gets the confidence $\false$,
the {\bf false} confidence level with respect to truth ordering, $\bot_t$
(see Section \ref{lattice}). 
Each success node labeled $(r, \theta)$, where $r$ is a 
rule in $P$, and $c$ is the confidence of rule $r$, 
gets the confidence $c$. \\ 
2. Suppose $u$ is a rule node labeled $(r, \theta)$, 
such that the confidence of $r$ is $c$,
its (conjunctive) mode is $\mu_r$, and  
the confidences of the children of $u$ are $c_1, \ldots, c_m$. 
Then $u$ gets 
the confidence $c \andmode c_1 \andmode \cdots \andmode c_m$. \\ 
3. Suppose $u$ is a goal node labeled $A$, 
with a (disjunctive) mode $\mu_p$ such that 
the confidences of its children are $c_1, \ldots, c_k$. Then $u$ gets the 
confidence $c_1 \ormode \cdots \ormode c_k$. 
\end{defn}

We recall the notions of identity and annihilator from algebra (\eg see 
Ullman \cite{ull-89-bk-vol2-db+kbs}). 
Let $c \in \conflat$ be any element of the confidence lattice and 
$\odot$ be any 
operation of the form $\wedge_\mu$ or of the form $\vee_\mu$, $\mu$ being any 
of the modes discussed in Section \ref{prob-calc}. Then $c$ is an 
{\em identity} with respect to $\odot$, 
if $\forall d \in \conflat,~c \odot d~=~d \odot 
c~=~d$. It is an {\em annihilator} with respect to $\odot$, 
if $\forall d \in \conflat,~
c \odot d~=~d \odot c~=~c$. 
The proof of the following proposition is straightforward. 

\begin{prop} \label{identity-annihilator} 
The truth-value $\bot_t = \false$ is an identity with respect to disjunction 
and an annihilator with respect to 
conjunction. The truth-value $\top_t = \true$ is an identity with respect to 
conjunction and an annihilator with respect to disjunction. 
These claims hold for all modes discussed in 
Section \ref{prob-calc}. 
\end{prop}

In view of this proposition, we can consider only DPTs without failure 
nodes without losing any generality. 

We now proceed to prove the soundness and completeness theorems.
First, we need some definitions.

\begin{defn}
A {\em branch} $B$ of a DPT $T$ is a set of nodes of $T$, defined as follows. 
The root of $T$ belongs to $B$. Whenever a goal node is in $B$, exactly 
one of its rule children (if any) belongs to $B$. Finally, whenever a rule 
node belongs to $B$, all its goal children belong to $B$. 
We extend this definition to the subtrees of a DPT in the obvious way.
A {\em subbranch} of $T$ rooted at a goal node $G$
is the branch of the subtree of $T$ rooted at $G$.

We can associate 
a substitution with a (sub)branch $B$ as follows. 
(1) The substitution associated 
with a success node labeled $(r, \theta)$ is just $\theta$. 
(2) The 
substitution associated with an internal goal node is simply the 
substitution associated with its unique rule child in $B$. 
(3) The substitution 
associated with an internal rule node $u$ in $B$ which is labeled $(r, \theta)$ 
is the composition of $\theta$ and the substitutions associated 
with the goal children of $u$. 

The substitution  associated with a branch is that associated with its root.
\end{defn}

We say a DPT $T$ is {\em well-formed} if it satisfies the following conditions:
(i) $T$ is proper,
(ii) For every goal node $G$ in $T$,
for any two (sub)branches $B_1, B_2$ of $T$ rooted at $G$,
the substitutions associated with $B_1$ and $B_2$ are distinct.

The second condition ensures no two branches correspond to the same 
classical ``proof" of the goal or a sub-goal.
Without this condition, since the probabilistic conjunctions and
disjunctions are not idempotent, the confidence of the same proof could be
wrongly combined giving an incorrect confidence for the (root of the) DPT.

Henceforth, we will only consider well-formed DPTs,
namely, DPTs that are proper, have no failure nodes, and have
distinct substitutions for all (sub) branches corresponding
to a goal node, for all goal nodes.

\begin{thm} [Soundness] \label{soundness}
Let $P$ be a p-program and $G$ a (ground) goal. 
If there is a finite well-formed DPT for $G$ with respect to $P$ 
with an associated confidence $c$ at its root, then $c \leq_t lfp(T_P)(G)$. 
\end{thm}

\noindent
{\bf Proof.}
First, we make the following observations regarding the combination functions
of Theorem \ref{thm:combinations}:

\noindent
(1) Conjunctive combination functions (all modes) are monotone. \\
(2) Disjunctive combination functions (all modes) are monotone. \\
(3) If $F$ and $G$ are confidence levels, then
$\cf (F \wedge_{\mu} G) \le_t \cf (F)$ and
$\cf (F) \le_t \cf (F \vee_{\mu} G)$
for all conjunctive and disjunctive combination functions (all modes).

\vs
We prove the soundness theorem 
by induction on the height of the DPT.
Assume the well-formed DPT $T$ of height $h$ is for the goal $G$.
Note that $T$ has an odd height, $h = 2k -1$ for some $k \ge 1$,
since it is a proper DPT with no failure nodes
(see Remark 7 at the beginning of this section).

\vs
\noindent
\underline{Basis}: $k = 1$.
In this case the DPT consists of the goal root labeled $G$
and one child labeled $(r,\theta)$,
where $r$ is a rule in $P$ whose head is unifiable with $G$.
Note that this child node is a success leaf.
\ie\ it represents a fact.
Obviously, in the first iteration of $T_P$,
$\cf (G) = c_r$,
where $c_r$ is the confidence level of
$r$.
It follows from
the monotonicity of $T_P$, that
$c = c_r \leq_t lfp(T_P)(G)$.

\vs
\noindent
\underline{Induction}: $k > 1$.
Assume the inductive hypothesis holds for every DPT of height $h' = 2k' -1$,
where $k' < k$. Consider the DPT $T$ for $G$. 
The root $G$ has rule children $R_i$ labeled $(r_i, \theta_i)$.
Each $R_i$ is either a fact, or has goal children
$G_{i_1}, \ldots, G_{i_{n_i}}$.
Consider the subtrees of $T$ rooted at these goal grand children of $G$.
By the inductive hypothesis, the confidence levels $c_{i_j}$ associated with
the goal grand children $G_{i_j}$ by the DPT are less than or equal to their
confidence levels calculated by $T_P$, \ie,
$c_{i_j} \leq_t lfp(T_P)(G_{i_j})$.
Hence, by properties (1)-(3) above, the confidence level associated 
to $G$ by $T$ is less than or equal to the confidence level of $G$
obtained by another application of $T_P$, $c \leq_t T_P(lfp(T_P))(G)$.
Hence $c \leq_t lfp(T_P)(G)$. 
Note that $T$ must be well-formed otherwise this argument is not valid.
\qed

\begin{thm} [Completeness] \label{completeness} 
Let $P$ be a p-program and $G$ a goal such that for some number $k < \omega$, 
$lfp(T_P)(G) = T_P\uparrow k(G)$. Then there is a finite DPT $T$ for 
$G$ with respect to $P$ with an associated confidence $c$ at its root, 
such that 
$lfp(T_P)(G) \leq_t c$. 
\end{thm}

\noindent
{\bf Proof.}
Let $k$ be the
smallest number such that $T_P\uparrow k (G) = lfp(T_P)(G)$.
We shall show by induction on $k$ that there is a DPT
$T$ for $G$ with respect to $P$ such that the confidence computed by it is at
least $lfp(T_P)(G)$. 
 
\vs
\noindent
\underline{Basis}: 
$k = 0$. This implies $lfp(T_P)(G) = \false$. This case is trivial.
The DPT consists of a failure node labeled $G$.
 
\vs
\noindent
\underline{Induction}: 
Suppose the result holds for a certain number $n$. 
We show that it also holds for $n+1$.
Suppose $A$ is a ground atom such that $lfp(T_P)(A) = c =
T_P\uparrow n+1 (A)$. 
Now, 
%
$T_P\uparrow n+1 (A) =
\vee_{\mu_p}~\{c_r \wedge_{\mu_r} 
T_P\uparrow n (B_1) \wedge_{\mu_r} \cdots \wedge_{\mu_r} 
T_P\uparrow n (B_m) | $
There exists a rule $r$ such that $\mu_r$ is the mode associated with its
body, and $\mu_p$ is the mode associated with its head, and 
there exists a ground
substitution $\theta$ such that $r\theta \equiv A \shortprulearg{c_r} B_1,
\ldots, B_m\}$. 

Consider the DPT for $A$ obtained as follows. Let
the root be labeled $A$. The root has a rule child corresponding to each
rule instance used in the above computation of $T_P\uparrow n+1 (A)$. Let
$v$ be a rule child created at this step, and suppose
$A \shortprulearg{c_r} B_1, \ldots, B_m$ is the rule instance corresponding
to it and let $\theta$ be the substitution used to unify the head of the
original rule with the atom $A$. Then $v$ has $m$ goal children with
labels $B_1, \ldots, B_m$ respectively. Finally, by induction hypothesis,
we can assume that 
(i) a DPT for $B_i$ is rooted at the node labeled $B_i$, and 
(ii) the confidence computed by this latter tree is
at least 
$lfp(T_P)(B_i) = T_P\uparrow n (B_i)$, $ 1 \leq i \leq m$. In this
case, it readily follows from the definition of the confidence computed by
a proof-tree that the confidence computed by $T$ is at least
$\vee_{\mu_p}~\{c_r \wedge_{\mu_r} conf(body(r)) | r$ is a rule defining
$A$, $c_r$ is the confidence associated with it, $\mu_r$ is the mode
associated with its head, and $\mu_p$ is the mode associated with its
body$\}$. 

But this confidence is exactly $T_P\uparrow n+1 (A)$. The induction
is complete and the theorem follows.
\qed

\vs
Theorems~\ref{soundness} and~\ref{completeness} together show that the 
confidence of an arbitrary ground atom computed according to the 
fixpoint semantics and using an appropriate
disjunctive proof tree is the same.
This in turn is the same 
as the confidence associated according to the (valuation based) declarative 
semantics. 
In particular, as we will discuss in Section \ref{termination},
when the disjunctive mode associated with all recursive predicates
is positive correlation, the theorems guarantee that the exact confidence 
associated with the goal can be effectively computed by constructing 
an appropriate finite DPT 
(according to Theorem \ref{completeness})
for it. Even when these modes {\em are} used indiscriminately, we can still 
obtain the confidence associated with the goal with an arbitrarily high 
degree of accuracy, by constructing DPTs of appropriate height.


\section{Termination and Complexity} \label{termination} 

In this section, we first compare our work with that of Ng and Subrahmanian 
\cite{Ng:PLP,NgSub90:SFforSS+CP-in-DD} (see Section \ref{intro} for a general comparison with non-probabilistic 
frameworks). First, let us examine the (only) ``mode" for disjunction 
used by them\footnote{Their framework allows an infinite class of ``conjunctive 
modes". Also, recall  they represent only beliefs.}. They combine the 
confidences of an atom $A$ coming from different derivations by taking their 
intersection. Indeed, the bottom of their lattice is a valuation (called  
``formula function" there) that assigns the interval $[0,1]$ to every atom. 
From the trilattice structure, it is clear that (i) their bottom 
corresponds to $\bot_p$, and (ii) their disjunctive mode corresponds to 
$\joinp$.

\begin{ex} \label{problem1} 
\hsp $r_1$: $p(X, Y): [V_1\times V_3, V_2\times V_4] \la e(X, Z): [V_1, V_2], 
~~$$p(Z, Y): [V_3, V_4]$. \\ 
\hsp $r_2$: $p(X, Y): [V_1, V_2] \la e(X, Y): [V_1, V_2]$. \\ 
\hsp $r_3$: $e(1, 2): [1, 1]$. \\
\hsp $r_4$: $e(1, 3): [1, 1]$. \\
\hsp $r_5$: $e(3, 2): [0.9, 0.9]$.

\vs
This is a pf-program in the framework of Ng and Subrahmanian 
\cite{NgSub90:SFforSS+CP-in-DD}.
In a pf-rule each literal is annotated by an interval
representing the lower-bound and upper-bound of belief\footnote{
To be precise, each {\em basic formula}, which is a conjunction
or a disjunction of atoms, is annotated.}.
Variables can appear in the annotations,
and the annotation of the head predicate is usually a function of
body literals annotations.
The program in this example is basically the transitive closure
program, with independence as the conjunctive mode in the first rule.
The disjunctive function for the $p$ predicate,
as explained above, is interval intersection.
Let us denote the operator $T_P$ defined by them as
$\tpann$ for distinguishing 
it from ours. 
It is not hard to see that this program is inconsistent in their
framework, and $lfp(\tpann)$would assign an empty probability 
range for $p(1, 2)$. This is due to the existence of two derivations
for $p(1, 2)$, with non-overlapping intervals. This is
quite unintuitive. Indeed, there is a definite path (with probability 1)
corresponding to the edge $e(1, 2)$.
One may wonder whether it makes sense to compare 
this approach with ours on an example program which is 
inconsistent according to their semantics. The point is that in this 
example, there is a certain path with probability [1,1] from 1 to 2, 
and an approach that regards this program as inconsistent is not 
quite intuitive.

Now, consider the p-program corresponding to the annotated program
$\{r_1, \ldots,$ $r_5\}$,
obtained by stripping off atom annotations in $r_1, r_2$ and shifting the
annotations in $r_3, \ldots, r_5$ to the associated rules. Also, associate
the confidence level $\true$ with $r_1, r_2$. For uniformity and ease
of comparison, assume the doubt ranges are all $[0,0]$. As an example, let
the conjunctive mode used in $r_1, r_2$ be independence and let the
disjunctive mode used be positive correlation (or, in this case,
even ignorance!).
Then $lfp(T_P)$ would
assign the confidence $\langle[1,1],~[0,0]\rangle$ to $p(1,2)$,
which agrees with our intuition.
Our point, however, is
{\sf not} that intersection is a ``wrong" mode. 
Rather, we stress that
different combination rules (modes) are appropriate for different
situations.
\qed
\end{ex}

\begin{ex} \label{problem2}
Now consider the following pf-program
($r_1$ and $r_2$ are the same as previous example):

\noindent
\hsp $r_1$: $p(X, Y): [V_1\times V_3, V_2\times V_4] \la 
e(X, Z): [V_1, V_2], ~~$$p(Z, Y): [V_3, V_4]$. \\
\hsp $r_2$: $p(X, Y): [V_1, V_2] \la e(X, Y): [V_1, V_2]$. \\
\hsp $r_6$: $e(1, 2): [0, 1]$. \\
\hsp $r_7$: $e(1, 1): [0, 0.9]$. 

\vs
In this case, the least fixpoint of $\tpann$ is only 
attained at $\omega$ and it assigns the range $[0, 0]$ to 
$p(1,1)$ and $p(1, 2)$. 
Again, the result is unintuitive for this example. Since $\tpann$ is 
not continuous, one can easily write programs such that no reasonable 
approximation to $lfp(\tpann)$ can be obtained by iterating $\tpann$ an 
arbitrary (finite) number of times. (\bigeg, consider the program obtained 
by adding the rule $r_8$: $q(X,Y): [1,1]~\la~p(X,Y): [0,0]$ 
to $\{r_1,r_2,r_6,r_7\}$.) 
Notice that as long as one uses 
any arithmetic annotation function such that the probability of the 
head is less than the probability of the subgoals of $r_1$ (which is 
a reasonable annotation function), 
this problem will arise. 
The problem (for the unintuitive behavior) 
lies with the mode for disjunction.
Again, we emphasize that different combination rules (modes) are
appropriate for different situations.

Now, consider the p-program corresponding to the annotated program 
$\{r_1, r_2, r_6, r_7\}$,
obtained in the same way as was done in Example \ref{problem1}.
Let the conjunctive mode used in $r_1, r_2$ be independence and let the 
disjunctive mode be positive correlation or ignorance.
Then $lfp(T_P)$ would assign the confidence 
level $\langle[0,1]~[0,0]\rangle$ to $p(1,2)$. This again agrees with our 
intuition. As a last example, suppose we start with the confidence 
$\langle[0,0.1],~[0,0]\rangle$ for $e(1,2)$ instead.
Then under positive correlation (for disjunction)
$lfp(T_P)(p(1,2)) = \langle[0,0.1],~[0,0]\rangle$,
while ignorance leads to $lfp(T_P)(p(1,2)) = \langle[0,1],~[0,0]\rangle$. 
The former makes more intuitive sense, although the latter (more conservative 
under $\leq_p$) is obviously 
{\sf not} wrong. Also, in the latter case, the $lfp$ is reached only 
at $\omega$. 
\qed
\end{ex}

Now, we discuss termination and complexity issues of p-programs.
Let the {\em closure ordinal} of $T_P$ 
be the smallest ordinal $\alpha$ such that 
$T_P\uparrow \alpha = lfp(T_P)$.
We have the following

\begin{fact}
\label{thm:ordinal}
Let $P$ be any p-program. Then 
the closure ordinal of $T_P$ can be as high as
$\omega$ but no more.
\end{fact} 

\noindent
{\bf Proof.}
The last p-program discussed in Example \ref{problem2}
has a closure ordinal of $\omega$.
Since $T_P$ is continuous (Theorem \ref{thm:tpmonotone})
its closure ordinal is at most $\omega$.
\qed

\begin{defn}
\label{defn:dataComplexity}
({\em Data Complexity})
We define the {\em data complexity} \cite{Va85:QLD} of a p-program $P$ 
as the time complexity of computing the least fixpoint of $T_P$ as a
function of the 
size of the database, \ie the number of constants occurring in 
$P$\footnote{With many rule-based systems with uncertainty, we cannot always 
separate EDB and IDB predicates, which explains this slightly modified 
definition of data complexity.}. 
\end{defn}

It is well known that the data complexity for datalog programs is polynomial. 
An important question concerning any extension of DDBs to handle uncertainty 
is whether the data complexity is increased compared to datalog. We can show 
that under suitable restrictions (see below) the data complexity of 
p-programs is polynomial time. However, the proof cannot be obtained by 
(straightforward extensions of) the classical argument for the data 
complexity for datalog. In the classical case, once a ground atom is derived 
during bottom-up evaluation, future derivations of it can be ignored. In 
programming with uncertainty, complications arise because we {\em cannot} 
ignore alternate derivations of the 
same atom: the confidences obtained from them need to be combined, 
reinforcing the overall confidence of the atom. This calls for a new proof 
technique. 
Our technique makes use of the following additional notions.

Define a {\sl disjunctive derivation tree} (DDT) 
to be a
well-formed DPT (see Section \ref{proof-theory} for a definition) 
such that every goal and every substitution
labeling any node in the tree is ground.
Note that the height of a DDT with no failure nodes is an odd number
(see Remark 7 at the beginning of Section \ref{proof-theory}).
We have the following results.

\begin{prop}\label{ddt-bu-eval}
Let $P$ be a p-program and $A$ any ground atom in $B_P$.
Suppose the confidence
determined for $A$ in iteration $k \geq 1$ of bottom-up evaluation is $c$. 
Then there
exists a DDT $T$ of height $2k - 1$ for $A$
such that the confidence associated 
with $A$ by $T$ is exactly $c$.
\end{prop}

\noindent
{\bf Proof.}
The proof is by induction on $k$.

\vs
\noindent
\underline{Basis}: $k = 1$.
In iteration 1, bottom-up evaluation essentially
collects together all edb facts (involving ground atoms) and determines their
confidences from the program. Without loss of generality, we may suppose
there is at most one edb fact in $P$ corresponding to each
ground atom (involving an edb predicate).
Let $A$ be any ground atom whose confidence is determined
to be $c$ in iteration 1. Then there is an edb fact $r: \shortstdfact$
in $P$.
The associated DDT for $A$ corresponding to this iteration is the tree with
root labeled $A$ and a rule child labeled $r$. Clearly, the confidence
associated with the root of this tree is $c$,
and the height of this tree is $1$ ( $= 2k - 1$, for $k=1)$.
 
\vs
\noindent
\underline{Induction}: 
Assume the result for all ground atoms whose
confidences are determined (possibly revised) in iteration $k$. 
Suppose $A$ is a
ground atom whose confidence is determined to be $c$ in iteration $k+1$.
This implies there exist ground instances of rules
$r_1:~A \shortprulearg{\crone} B_1, \ldots, B_m; ~\muone, \mu_A$, 
$\cdots$; 
$r_k:~A \shortprulearg{\crk} C_1, \ldots, C_n; ~\muk, \mu_A$
such that
(i) the confidence of $B_i$ $, \ldots, C_j$
computed at the end of iteration $k$ is $c_i$
($, \ldots, d_j$), and (ii) $c = (\crone \wedge_{\muone}$
$(\bigwedge_{\muone}~\{c_1, \ldots, c_m\})$ $\orA \cdots \orA$
$(\crk \wedge_{\muk} (\bigwedge_{\muk}~\{d_1, \ldots, d_n\})$,
where \orA\ is the disjunctive mode for the predicate $A$.
By induction hypothesis, 
there are DDTs $T_{B_1}, \ldots, T_{B_m}$, $\ldots,
T_{C_1}, \ldots, T_{C_n}$,
each of height $2k -1$ or less, for the atoms
$B_1, \ldots, B_m$, $\ldots, C_1, \ldots, C_n$ which
exactly compute the confidences $c_1, \ldots, c_m$, $\ldots, d_1, \ldots, d_n$
respectively, corresponding to iteration $k$.
Consider the tree $T_{k+1}$ for $A$
by (i) making $r_1, \ldots, r_k$ rule children of
the root and (ii) making the $T_{B_1}, \ldots, T_{B_m}$, ($, \ldots,
T_{C_1}, \ldots, T_{C_n}$) subtrees of $r_1$ ($, \ldots, r_k$).
It is trivial to
see that $T_{k+1}$ is a DDT for $A$ and
its height is $2 + 2k -1 = 2(k+1) -1$.
Further the confidence $T_{k+1}$ computes for the
root $A$ is exactly $(\crone \wedge_{\muone}$
$(\bigwedge_{\muone}~\{c_1, \ldots, c_m\})$ $\orA \cdots \orA$
$(\crk \wedge_{\muk} (\bigwedge_{\muk}~\{d_1, \ldots, d_n\})$.
This completes the induction and the proof.
\qed
 
\vs
Proposition \ref{ddt-bu-eval} shows each iteration of bottom-up
evaluation corresponds in an essential manner to the construction of a set
of DDTs one for each distinct ground atom whose confidence is determined
(or revised) in that iteration. Our next objective is to establish a
termination bound on bottom-up evaluation.

\begin{defn}[Branches in DDTs] 
DDT branches are defined similar to those of DPT.
Let $T$ be a DDT. Then a branch of $T$ is a subtree of $T$, 
defined as follows. 

\noindent
(i) The root belongs to every branch. \\
(ii) whenever a goal node belongs to a branch,
exactly one of its rule children, 
belongs to the branch. \\
(iii) whenever a rule node belongs to a branch,
all its goal children belong 
to the branch.
\end{defn}

\begin{defn}[Simple DDTs] 
Let $A$ be a ground atom and $T$ any DDT (not necessarily for $A$). 
Then $T$ is $A$-{\sf non-simple} provided it has a branch 
containing two goal nodes 
$u$ and $v$ such that $u$ is an ancestor of $v$ and 
both are labeled by atom $A$. 
A DDT is $A$-{\sf simple} if it is not $A$-non-simple. Finally, a DDT is 
{\sf simple} if it is $A$-simple for every atom $A$. 
\end{defn} 

Let $T$ be a DDT and $B_i$ be a branch of $T$ in which an atom $A$
appears.
Then we define the {\em number of violations of simplicity} of $B_i$ 
with respect to $A$ 
to be one less than the total number of times 
the atom $A$ occurs in $B_i$. 
The number of violations of the simplicity of the DDT $T$
with respect to $A$ is the sum of the number of violations
of the branches of $T$ in which $A$ occurs.
Clearly, $T$ is $A$-simple exactly when the number of violations
with respect to $A$ is 0. Our first major result of this section follows. 

\begin{thm}\label{term-bound}
Let $P$ be a p-program such that only positive correlation is used as the
disjunctive mode for recursive predicates. Let $max\{height(T_A) \mid $A$ \in
B_P$, and $T_A$ is any simple DDT for $A\}$ = $2k - 1$, $k \geq 1$.
Then at most $k+1$ iterations of naive bottom-up evaluation are needed to
compute the least fixpoint of $T_P$.
\end{thm}

Essentially, for p-programs $P$ satisfying the conditions mentioned above, 
the theorem (i) shows that naive bottom-up evaluation of 
$P$ is guaranteed to terminate, and 
(ii) establishes an upper bound on the number of iterations 
of bottom-up evaluation for computing the least fixpoint, 
in terms of the maximum height of any simple tree for any 
ground atom. This is the first step in showing that p-programs of this 
type have a polynomial time data complexity. 
We will use the next three lemmas 
(Lemmas~\ref{lem:key}--\ref{lem:boundForA}) in proving this 
theorem.

\begin{lemma} 
\label{lem:key}
Let $A \in B_{P}$ be any ground atom, and let 
$T$ be a DDT for $A$ corresponding to $T_{P}\uparrow l$, for some $l$. 
Suppose $T$ is $B$-non-simple, for some atom $B$. 
Then there is a DDT $T'$ for $A$ 
with the following properties: 

\noindent
(i) the certainty of $A$ computed by $T'$ equals that computed by $T$. \\
(ii) the number of violations of simplicity of $T'$ with respect to $B$ 
is less than that of $T$. 
\end{lemma}

\noindent
{\bf Proof}.
Let $T$ be the DDT described in the hypothesis of the claim. Let $A$ be 
the label of the root $u$ of $T$,
and assume without loss of generality that $B$ is identical to $A$. 
(The case when $B$ is distinct from $A$ is similar.)
Let $v$ be the last goal node from the root down
(e.g. in the level-order), which is distinct from the 
root and is labeled by $A$. 
Since $T$ corresponds to applications of the $T_P$ operator, 
we have the following. 
 
\noindent
(*) Every branch of $v$ must be isomorphic to some branch of 
$u$ which does not contain the node $v$. 

This can be seen as follows. Let 
$p < l$ be the iteration such that the subtree of $T$ rooted at $v$, 
say $T_v$,  
corresponds to $T_{P}\uparrow p$. Then clearly, every rule applicable 
in iteration $p$ is also applicable in iteration $k$.
This means every branch 
of $T_v$ constructed from a sequence of rule applications is also 
constructible in iteration $k$ and hence
there must be a branch of $T$ that is 
isomorphic to such a branch. It follows from the isomorphism that 
the isomorphic branch of $T$ cannot contain the node $v$. 

Associate a logical formula with each node of $T$ as follows. 

\noindent
(i) The formula associated with each (rule) leaf is {\bf true}.  \\
(ii) The formula associated with a goal node with rule children 
$r_1, \ldots, r_m$ and associated formulas 
$F_1, \ldots, F_m$, is $F_1 \vee \cdots \vee F_m$.  \\
(iii) The formula associated with a rule parent 
with goal children $g_1, \ldots, g_q$
and associated formulas $G_1, \ldots, G_q$ is 
$G_1 \wedge \cdots \wedge G_q$. 

Let the formula associated with the node $v$ be $F$.
To simplify the exposition, but at no loss of generality, let us assume that 
in $T$, every goal node has exactly two rule children. Then 
the formula associated with 
the root $u$ can be expressed as $E_1 \vee (E_2 \wedge (E_3 \vee 
(\cdots E_{s-1} \wedge (E_s \vee F)) \cdots ))$.

By (*) above, we can see that 
$F$ logically implies $E_1$, $F \rimplies E_1$. 
By the structure of a DDT, 
we can then express $E_1$ as $(F \vee G)$, for some formula $G$. 
Construct a DDT $T'$ 
from $T$ by deleting the parent of the node $v$, 
as well as the subtree rooted 
at $v$. We claim that 

\noindent
(**) The formula associated with the root of $T'$ is equivalent to that 
     associated with the root of $T$. 

To see this, notice that the formula associated with the root of $T$ can now 
be expressed as $(F \vee G) \vee (E2 \wedge (E3 \vee 
(\cdots Es-1 \wedge (Es \vee F)) \cdots ))$. By simple application of 
propositional identities, it can be seen that this formula is equivalent 
to $(F \vee G) \vee (E2 \wedge (E3 \vee (\cdots Es-1 \wedge (Es)) \cdots ))$. 
But this is exactly the formula associated with the root of T', which 
proves (**).

Finally, we shall show that $\vee_{pc}$, together with any conjunctive mode, 
satisfy the following absorption laws: 

\noindent
\hsp $a \vee_{pc} (a \wedge_{\mu} b) = a$. \\
\hsp $a \wedge_{pc} (a \vee_{\mu} b) = a$.

The first of these laws follows from the fact that for all modes $\mu$ 
we consider in this paper, $(a \wedge_{\mu} b) \leq_t a$,
where $\leq_t$ is the 
lattice ordering. The second is the dual of the first. 

In view of the absorption laws, it can be seen that the certainty for $A$ 
computed by $T'$ above is identical to that computed by $T$. 
This proves the lemma, since $T'$ has at least one fewer violations of 
simplicity with respect to $A$. \qed

\begin{lemma}
\label{lem:simple}
Let $T$ be a DDT for an atom $A$. 
Then there is a simple 
DDT for $A$ such that the certainty of $A$ computed by it is identical to 
that computed by $T$. 
\end{lemma}

\noindent
{\bf Proof}.
Follows by an inductive argument using Lemma \ref{lem:key}. \qed 

\begin{lemma}
\label{lem:boundForA}
Let $A$ be an atom and $2h-1$ be the maximum height of any simple 
DDT for $A$. Then certainty of $A$ in $T_{P}\uparrow l$ 
is identical to 
that in $T_{P}\uparrow h$, for all $l \geq h$. 
\end{lemma}

\noindent
{\bf Proof}.
Let $T$ be the DDT for $A$ corresponding to $T_{P}\uparrow l$.
Note that height($T) = 2l-1$.
Let $c$ represent the certainty computed by $T$ for $A$,
which is $c = T_{P}\uparrow l(A)$.
By Lemma \ref{lem:simple},
there is a simple DDT, say $T'$, for $A$, which computes the same 
certainty for $A$ as $T$. Clearly,
height($T') \le 2h-1$. 
Let $c'$ represent the certainty computed by $T'$ for $A$,
$c = c'$.
By the soundness theorem, and monotonicity of $T_P$,
we can write
$c' \leq T_P \uparrow h (A) \leq T_P \uparrow l (A) = c$.
It follows that 
$T_{P}\uparrow l (A) = T_{P}\uparrow h (A)$.
\qed 

\vsp
Now we can complete the proof of Theorem \ref{term-bound}.

\noindent
{\bf Proof of Theorem \ref{term-bound}}.
Let $2k-1$ be the maximum height of any simple DDT for any atom. 
It follows from the above Lemmas that the certainty of any atom 
in $T_{P}\uparrow l$ is identical to that in $T_{P}\uparrow k$, 
for all $l \geq k$, from which the theorem follows. \qed

It can be shown that the height of simple DDTs is polynomially bounded by the 
database size. This makes the above result significant. 
This allows us to prove the following theorem regarding the data complexity 
of the above class of p-programs.  

\begin{thm} \label{complexity} 
Let $P$ be a p-program such that only positive correlation is used as the 
disjunctive mode for recursive predicates. 
Then its least fixpoint can be computed in time 
polynomial in the database size. In particular, bottom-up naive evaluation 
terminates in time polynomial in the size of the database, yielding the 
least fixpoint. 
\end{thm} 

\noindent
{\bf Proof.} By Theorem \ref{term-bound} we know that the least fixpoint
model of $P$ can be computed in at most $k+1$ iterations where
$h = 2k - 1$ is the maximum height of any simple DDT for any ground
atom with respect to $P$
($k$ iterations to arrive at the fixpoint, and one extra iteration to verify 
that a fixpoint has been reached.)
Notice that each goal node in a DDT corresponds to a database predicate.
Let $K$ be the maximum arity of any predicate in $P$,
and $n$ be the number
of constants occurring in the program.
Notice that under the data complexity measure 
(Definition \ref{defn:dataComplexity}) $K$ is a constant.
The maximum number of distinct goal nodes that
can occur in any branch of a simple DDT is $n^K$. This implies the height
$h$ above is clearly a polynomial in the database size $n$.
We have thus shown
that bottom-up evaluation of the least fixpoint terminates in a polynomial
number of iterations. The fact that the amount of work done in each
iteration is polynomial in $n$ is easy to see. The theorem follows.
\qed

We remark that our proof of Theorem \ref{complexity} implies a similar 
result for van Emden's framework. To our knowledge, this is the first 
polynomial time result for rule-based programming with (probabilistic) 
uncertainty\footnote{It 
is straightforward to show that the data complexity for the framework of 
\cite{Ng:PLP} is polynomial, although the paper does not address this issue. 
However, that framework only allows constant annotations and is of limited 
expressive power.}. 
We should point out that the polynomial time complexity is preserved whenever 
modes other than positive correlation are associated with non-recursive 
predicates (for disjunction). More generally, suppose $R$ is the set of all 
recursive predicates and $N$ is the set of non-recursive predicates in the KB, 
which are 
possibly defined in terms of those in $R$. Then any modes can be freely used 
with the predicates in $N$ while keeping the data complexity polynomial. 
Finally, if we know that the data does not contain cycles, we can use any 
mode even with a recursive predicate and still have a polynomial time 
data complexity. 
We also note that the framework of annotation functions used in
\cite{KiSub92:GAPs} enables an infinite family of modes to be used in 
propagating confidences from rule bodies to heads. The major differences with 
our work are (i) in \cite{KiSub92:GAPs} a fixed ``mode" for disjunction is 
imposed unlike our framework, and (ii) they do not study the complexity 
of query answering, whereas we establish the conditions under which the 
important advantage of polynomial time data complexity of classical 
datalog can be retained. More importantly, our work has generated useful 
insights into how modes (for disjunction) affect the data complexity. 
Finally, a note about the use of positive correlation as the disjunctive 
mode for recursive predicates (when data might contain cycles). The rationale 
is that different derivations of such recursive atoms could involve 
some amount of overlap (the degree of overlap depends on the data). Now, 
positive correlation (for disjunction) tries to be conservative (and hence 
sound) by assuming 
the extent of overlap is maximal, so the combined confidence of the 
different derivations is the least possible (under $\leq_t$). Thus, it 
{\em does}  
make sense even from a practical point of view.


\section{Conclusions} \label{concl} 

We motivated the need for modeling both belief and doubt in a framework 
for manipulating uncertain facts and rules. We have developed a framework 
for probabilistic deductive databases, capable of manipulating both 
belief and doubt, expressed as probability intervals. 
Belief doubt pairs, called confidence levels, 
give rise to a rich algebraic structure called a trilattice. 
We developed a probabilistic calculus permitting different modes for combining 
confidence levels of events. We then developed the framework of p-programs 
for realizing probabilistic deductive databases. p-Programs inherit the 
ability to ``parameterize" the modes used for combining confidence 
levels, from our probabilistic calculus. We have developed a declarative 
semantics, a fixpoint semantics, and proved their equivalence. We have also 
provided  a sound and complete proof procedure for p-programs. 
We have shown that under 
disciplined use of modes, we can retain the important advantage of 
polynomial time 
data complexity of classical datalog, in this extended framework. We have also 
compared our framework with related work with respect to the aspects of 
termination and intuitive behavior (of the semantics). The parametric nature 
of modes in p-programs is shown to be a significant advantage 
with respect to these 
aspects. Also, the analysis of trilattices shows insightful relationships 
between previous work (\eg Ng and Subrahmanian 
\cite{Ng:PLP,NgSub90:SFforSS+CP-in-DD}) and ours. 
Interesting open issues which merit further research include (1) semantics 
of p-programs under various trilattice orders and various modes, including 
new ones, (2) query optimization, (3) handling inconsistency in a framework 
handling uncertainty, such as the one studied here.

\section*{Acknowledgments}

The authors would like to thank the anonymous referees 
for their careful reading and their comments,
many of which have resulted in significant improvements to the paper.


\begin{thebibliography}{}

\bibitem[\protect\citename{Abiteboul {\em et~al.}\relax,
  }1991]{abiteboul-etal-nulls-tcs91}
Abiteboul, S., Kanellakis, P., \& Grahne, G. (1991).
\newblock On the representation and querying of sets of possible worlds.
\newblock {\em Theoretical computer science}, {\bf 78}, 159--187.

\bibitem[\protect\citename{Baldwin, }1987]{baldwin-evidential-lp-jfuzzy87}
Baldwin, J.~F. (1987).
\newblock Evidential support logic programming.
\newblock {\em Journal of fuzzy sets and systems}, {\bf 24}, 1--26.

\bibitem[\protect\citename{Baldwin \& Monk,
  }1987]{baldwin-monk-evidential-fuzzy-lp-tr87}
Baldwin, J.~F., \& Monk, M. R.~M. (1987).
\newblock {\em Evidence theory, fuzzy logic, and logic programming}.
\newblock Tech. Report ITRC No. 109. University of Bristol, Bristol, UK.

\bibitem[\protect\citename{Barbara {\em et~al.}\relax, }1990]{Ba89:PRDM}
Barbara, D., Garcia-Molina, H., \& Porter, D. (1990).
\newblock A probabilistic relational data model.
\newblock {\em Pages  60--64 of:} {\em Proc. advancing database technology,
  {EDBT}, 90}.

\bibitem[\protect\citename{Barbara {\em et~al.}\relax, }1992]{BGP92}
Barbara, D., Garcia-Molina, H., \& Porter, D. (1992).
\newblock The management of probabilistic data.
\newblock {\em {IEEE} transactions on knowledge and data engineering}, {\bf
  4}(5), 487--502.

\bibitem[\protect\citename{Blair \& Subrahmanian, }1989a]{Bl89:ParaLP}
Blair, H.~A., \& Subrahmanian, V.~S. (1989a).
\newblock Paraconsistent foundations for logic programming.
\newblock {\em Journal of non-classical logic}, {\bf 5}(2), 45--73.

\bibitem[\protect\citename{Blair \& Subrahmanian,
  }1989b]{blair-vs-para-lp-tcs89}
Blair, H.~A., \& Subrahmanian, V.~S. (1989b).
\newblock Paraconsistent logic programming.
\newblock {\em Theoretical computer science}, {\bf 68}, 135--154.

\bibitem[\protect\citename{Boole, }1854]{boole-prob-logic-bk-1854}
Boole, G. (1854).
\newblock {\em The laws of thought}.
\newblock London: Mcmillan.

\bibitem[\protect\citename{Carnap, }1962]{carnap-prob-logic-bk62}
Carnap, R. (1962).
\newblock {\em The logical foundations of probability}.
\newblock University of Chicago Press.
\newblock 2nd. Edn.

\bibitem[\protect\citename{Debray \& Ramakrishnan,
  }1994]{debray-raghu-ghcp-jan94}
Debray, S., \& Ramakrishnan, R. (1994).
\newblock {\em Generalized {Horn} clause programs}.
\newblock manuscript.

\bibitem[\protect\citename{Dekhtyar \& Subrahmanian, }1997]{dekhtyarVS97}
Dekhtyar, A., \& Subrahmanian, V.~S. (1997).
\newblock Hybrid probabilistic program.
\newblock {\em Pages  391--405 of:} {\em Proc. 14th intl. conf. on logic
  programming}.

\bibitem[\protect\citename{Dong \& Lakshmanan,
  }1992]{dong-laks-92-jicslp-ddb-nulls}
Dong, F., \& Lakshmanan, L.V.S. (1992).
\newblock Deductive databases with incomplete information.
\newblock {\em Pages  303--317 of:} {\em Joint intl. conf. and symp. on logic
  programming}.
\newblock (extended version available as Tech. Report, Concordia University,
  Montreal, 1993).

\bibitem[\protect\citename{Fagin {\em et~al.}\relax,
  }1990]{fagin-etal-prob-logic-inc92}
Fagin, R., Halpern, J., \& Megiddo, N. (1990).
\newblock A logic for reasoning about probabilities.
\newblock {\em Information and computation}, {\bf 87}(1/2), 78--128.

\bibitem[\protect\citename{Fenstad, }1980]{Fe80:ProbFOL}
Fenstad, J.~E. (1980).
\newblock The structure of probabilities defined on first-order languages.
\newblock {\em Pages  251--262 of:} Jeffrey, R.~C. (ed), {\em Studies in
  inductive logic and probabilities, volume 2}.
\newblock University of California Press.

\bibitem[\protect\citename{Fitting, }1988]{Fi88:LPLat}
Fitting, M.~C. (1988).
\newblock Logic programming on a topological bilattice.
\newblock {\em Fundamenta informaticae}, {\bf 11}, 209--218.

\bibitem[\protect\citename{Fitting, }1991]{fitting-bilattice-lp-jlp91}
Fitting, M.~C. (1991).
\newblock Bilattices and the semantics of logic programming.
\newblock {\em Journal of logic programming}, {\bf 11}, 91--116.

\bibitem[\protect\citename{Fitting, }1995]{Fitting-letter}
Fitting, M.~C. 1995 (February).
\newblock Private Communication.

\bibitem[\protect\citename{{Frechet, M.},
  }1935]{frechet-ignorance-prob-bounds-fund-math35}
{Frechet, M.} (1935).
\newblock Generalizations du theoreme des probabilities totales.
\newblock {\em Fund. math.}, {\bf 25}, 379--387.

\bibitem[\protect\citename{Gaifman, }1964]{gaifman-prob-logic-isrjmath64}
Gaifman, H. (1964).
\newblock Concerning measures in first order calculi.
\newblock {\em Israel j. of math.}, {\bf 2}, 1--17.

\bibitem[\protect\citename{Ginsburg, }1988]{ginsburg-bilattice-compint88}
Ginsburg, M. (1988).
\newblock Multivalued logics: A uniform approach to reasoning in artificial
  intelligence.
\newblock {\em Computational intelligence}, {\bf 4}, 265--316.

\bibitem[\protect\citename{G{\"{u}}ntzer {\em et~al.}\relax,
  }1991]{guntzer-etal-newdir-unc-ddb-sigmod91}
G{\"{u}}ntzer, U., Kie{\ss}ling, W., \& Th{\"{o}}ne, H. (1991).
\newblock New directions for uncertainty reasoning in deductive databases.
\newblock {\em Pages  178--187 of:} {\em Proc. {ACM} {SIGMOD} intl. conf. on
  management of data}.

\bibitem[\protect\citename{Hailperin, }1984]{Hailperin84:ProbLog}
Hailperin, T. (1984).
\newblock Probability logic.
\newblock {\em Notre dame j. of formal logic}, {\bf 25}(3), 198--212.

\bibitem[\protect\citename{Halpern, }1990]{Hal-fol+prob-jAI-90}
Halpern, J.~Y. (1990).
\newblock An analysis of first-order logics of probability.
\newblock {\em Journal of {AI}}, {\bf 46}, 311--350.

\bibitem[\protect\citename{Kifer \& Li, }1988]{Ki88:SemESUnc}
Kifer, M., \& Li, A. (1988).
\newblock On the semantics of rule-based expert systems with uncertainty.
\newblock {\em Pages  102--117 of:} Gyssens, M., Paradaens, J., \& van Gucht,
  D. (eds), {\em 2nd intl. conf. on database theory}.
\newblock Bruges, Belgium: Springer-Verlag LNCS-326.

\bibitem[\protect\citename{Kifer \& Lozinskii,
  }1989]{kifer-loz-logic-incons-lics89}
Kifer, M., \& Lozinskii, E.~L. (1989).
\newblock A logic for reasoning with inconsistency.
\newblock {\em Pages  253--262 of:} {\em Proc. 4th {IEEE} symp. on logic in
  computer science ({LICS})}.
\newblock Asilomar, CA: {IEEE} Computer Press.

\bibitem[\protect\citename{Kifer \& Lozinskii,
  }1992]{kifer-loz-logic-incons-jar}
Kifer, M., \& Lozinskii, E.~L. (1992).
\newblock A logic for reasoning with inconsistency.
\newblock {\em Journal of automated reasoning}, {\bf 9}(2), 179--215.

\bibitem[\protect\citename{Kifer \& Subrahmanian, }1992]{KiSub92:GAPs}
Kifer, M., \& Subrahmanian, V.~S. (1992).
\newblock Theory of generalized annotated logic programming and its
  applications.
\newblock {\em Journal of logic programming}, {\bf 12}, 335--367.

\bibitem[\protect\citename{Lakshmanan, }1994]{laks-epistemic-fsttcs94}
Lakshmanan, L. V.~S. (1994).
\newblock An epistemic foundation for logic programming with uncertainty.
\newblock  {\em Proc. intl. conf. on foundations of software technology and
  theoretical computer science}.
\newblock Madras, India: Springer Verlag.
\newblock Lecture Notes in Computer Science, vol. 880.

\bibitem[\protect\citename{Lakshmanan \& Sadri, }1994a]{LS:DIST}
Lakshmanan, L. V.~S., \& Sadri, F. (1994a).
\newblock Modeling uncertainty in deductive databases.
\newblock {\em Pages  724--733 of:} {\em Proc. intl. conf. on database and
  expert systems applications ({DEXA} '94)}.
\newblock Athens, Greece: Springer-Verlag, LNCS-856.

\bibitem[\protect\citename{Lakshmanan \& Sadri, }1994b]{laks-sadri-pddb-ilps94}
Lakshmanan, L. V.~S., \& Sadri, F. (1994b).
\newblock Probabilistic deductive databases.
\newblock {\em Pages  254--268 of:} {\em Proc. intl. logic programming
  symposium}.
\newblock Ithaca, NY: MIT Press.

\bibitem[\protect\citename{Lakshmanan \& Sadri, }1997]{Lakshmanan-Sadri:is97}
Lakshmanan, L. V.~S., \& Sadri, F. (1997).
\newblock Uncertain deductive databases: A hybrid approach.
\newblock {\em \infosystems}, {\bf 22}(8), 483--508.

\bibitem[\protect\citename{Lakshmanan \& Shiri, }1997]{tkde97}
Lakshmanan, L. V.~S., \& Shiri, N. (1997).
\newblock A parametric approach to deductive databases with uncertainty.
\newblock {\em Accepted to the {IEEE} transactions on knowledge and data
  engineering}.
\newblock (A preliminary version appeared in Proc. Intl. Workshop on Logic in
  Databases (LID'96), Springer-Verlag, LNCS-1154, San Miniato, Italy).

\bibitem[\protect\citename{Lakshmanan {\em et~al.}\relax, }{1997}]{probview96}
Lakshmanan, L. V.~S., {N. Leone}, {R. Ross}, \& {V. S. Subrahmanian}. ({1997}).
\newblock {ProbView: A Flexible Probabilistic Database System}.
\newblock {\em {{ACM} Transactions on Database Systems}}, {\bf 22}(3),
  419--469.

\bibitem[\protect\citename{Liu, }1990]{liu-ddb-nulls-ilps90}
Liu, Y. (1990).
\newblock Null values in definite programs.
\newblock {\em Pages  273--288 of:} {\em Proc. north american conf. on logic
  programming}.

\bibitem[\protect\citename{Ng, }1997]{Ng:tkde97}
Ng, R.~T. (1997).
\newblock Semantic, consistency, and query processing of empirical deductive
  databases.
\newblock {\em \tkde}, {\bf 9}(1), 32--495.

\bibitem[\protect\citename{Ng \& Subrahmanian, }1991]{NgSub:DS-theory-91}
Ng, R.~T., \& Subrahmanian, V.~S. (1991).
\newblock {\em Relating {Dempster-Shafer} theory to stable semantics}.
\newblock Tech. Report UMIACS-TR-91-49, CS-TR-2647. Institute for Advanced
  Computer Studies and Department of Computer Science University of Maryland,
  College Park, MD 20742.

\bibitem[\protect\citename{Ng \& Subrahmanian, }1992]{Ng:PLP}
Ng, R.~T., \& Subrahmanian, V.~S. (1992).
\newblock Probabilistic logic programming.
\newblock {\em Information and computation}, {\bf 101}(2), 150--201.

\bibitem[\protect\citename{Ng \& Subrahmanian, }1993]{NgSub90:SFforSS+CP-in-DD}
Ng, R.~T., \& Subrahmanian, V.~S. (1993).
\newblock A semantical framework for supporting subjective and conditional
  probabilities in deductive databases.
\newblock {\em Automated reasoning}, {\bf 10}(2), 191--235.

\bibitem[\protect\citename{Ng \& Subrahmanian, }1994]{NgSub90:SS-for-PLP}
Ng, R.~T., \& Subrahmanian, V.~S. (1994).
\newblock Stable semantics for probabilistic deductive databases.
\newblock {\em Information and computation}, {\bf 110}(1), 42--83.

\bibitem[\protect\citename{Nilsson, }1986]{nilsson-prob-logic-jai86}
Nilsson, N. (1986).
\newblock Probabilistic logic.
\newblock {\em {AI} journal}, {\bf 28}, 71--87.

\bibitem[\protect\citename{Sadri, }1991a]{Sad-IST-icde-90}
Sadri, F. (1991a).
\newblock Modeling uncertainty in databases.
\newblock {\em Pages  122--131 of:} {\em Proc. 7th {IEEE} intl. conf. on data
  engineering}.

\bibitem[\protect\citename{Sadri, }1991b]{Sa:RAQ}
Sadri, F. (1991b).
\newblock Reliability of answers to queries in relational databases.
\newblock {\em {IEEE} transactions on knowledge and data engineering}, {\bf
  3}(2), 245--251.

\bibitem[\protect\citename{Schmidt {\em et~al.}\relax,
  }1989]{schmidt-etal-qddb-dood89}
Schmidt, H., Steger, N., G{\"{u}}ntzer, U., Kie{\ss}ling, W., Azone, A., \&
  Bayer, R. (1989).
\newblock Combining deduction by certainty with the power of magic.
\newblock {\em Pages  103--122 of:} {\em Proc. 1st intl. conf. on deductive and
  object-oriented databases}.

\bibitem[\protect\citename{Steger {\em et~al.}\relax,
  }1989]{steger-etal-qddb-icde89}
Steger, N., Schmidt, H., G{\"{u}}ntzer, U., \& Kie{\ss}ling, W. (1989).
\newblock Semantics and efficient compilation for quantitative deductive
  databases.
\newblock {\em Pages  660--669 of:} {\em Proc. {IEEE} intl. conf. on data
  engineering}.

\bibitem[\protect\citename{Subrahmanian, }1987]{Sub87:QLP}
Subrahmanian, V.~S. (1987).
\newblock On the semantics of quantitative logic programs.
\newblock {\em Pages  173--182 of:} {\em Proc. 4th {IEEE} symposium on logic
  programming}.

\bibitem[\protect\citename{Ullman, }1989]{ull-89-bk-vol2-db+kbs}
Ullman, J.~D. (1989).
\newblock {\em Principles of database and knowledge-base systems}.
\newblock  Vol. II.
\newblock Maryland: Computer Science Press.

\bibitem[\protect\citename{van Emden, }1986]{vEm86:QuanDedFPTh}
van Emden, M.~H. (1986).
\newblock Quantitative deduction and its fixpoint theory.
\newblock {\em Journal of logic programming}, {\bf 4}(1), 37--53.

\bibitem[\protect\citename{{Vardi, M.Y.}, }1985]{Va85:QLD}
{Vardi, M.Y.} (1985).
\newblock Querying logical databases.
\newblock {\em Pages  57--65 of:} {\em Proc. 4th {ACM} {SIGACT}-{SIGMOD}
  symposium on principles of database systems}.

\end{thebibliography}
\end{document}